\documentclass[twocolumn,twocolappendix]{aastex631}

\usepackage{amsmath}

\newcommand{\jhrev}[1]{{\color{black}#1}}

\shorttitle{Analyzing Balmer Lines of BD and VLMS}
\shortauthors{Hashimoto et al.}

\graphicspath{{./}{figures/}}

\begin{document}

\title{Analyses of Multiple Balmer Emission Lines from Accreting Brown Dwarfs and Very Low Mass Stars}

\correspondingauthor{Jun Hashimoto}
\email{jun.hashimto@nao.ac.jp}

\author[0000-0002-3053-3575]{Jun Hashimoto}
\affil{Astrobiology Center, National Institutes of Natural Sciences, 2-21-1 Osawa, Mitaka, Tokyo 181-8588, Japan}
\affil{Subaru Telescope, National Astronomical Observatory of Japan, Mitaka, Tokyo 181-8588, Japan}
\affil{Department of Astronomy, School of Science, Graduate University for Advanced Studies (SOKENDAI), Mitaka, Tokyo 181-8588, Japan}

\author[0000-0003-0568-9225]{Yuhiko Aoyama}
\affil{School of Physics and Astronomy, Sun Yat-sen University, Guangdong 519082, People’s Republic of China}
\affil{Kavli Institute for Astronomy and Astrophysics, Peking University, Beijing 100084, People’s Republic of China}

\begin{abstract}
A planetary growth rate, a.k.a., the mass accretion rate, is a fundamental parameter in planet formation, as it determines a planet's final mass. Planetary mass accretion rates have been estimated using hydrogen lines, based on the models originally developed for accreting stars, known as the accretion flow model. Recently, Aoyama et al. (2018) introduced the accretion shock model as an alternative mechanism for hydrogen line emission. However, it remains unclear which model is more appropriate for accreting planets and substellar objects. To address this, we applied both models to archival data consisting of 96 data points from \jhrev{76} accreting brown dwarfs and very low-mass stars, with masses ranging from approximately 0.02 to 0.1 $M_\sun$, to test which model best explains their accreting properties. The results showed that the emission mechanisms of 15 data points are best explained by the shock model, while 55 data points are best explained by the flow model. For the 15 data points explained by the planetary shock model, the shock model estimates up to several times higher mass accretion rates than the flow model. As this trend is more pronounced for planetary mass objects, it is crucial to determine which emission mechanism is dominant in individual planets. We also discuss the physical parameters that determine the emission mechanisms and the variability of line ratios.
\end{abstract}


\section{Introduction} \label{sec:intro}

Planets form within protoplanetary disks, with the final phase of their formation controlled by the gas supplied from these disks \citep[e.g.,][]{tanigawa2002}. The planetary growth rate, a.k.a., the mass accretion rate, is a fundamental parameter in planet formation, as it determines the final mass of a planet. Traditionally, the accretion rate for T Tauri stars has been estimated using Balmer (at $\lambda \lesssim 364.6$ nm) and Paschen (at $\lambda \lesssim 820.4$ nm) continuum excesses \citep[e.g.,][]{Calvet1998ff,Gullbring1998TTS}. When most of the accretion energy is radiated as continuum excess, the luminosity of this excess corresponds to the accretion luminosity, which in turn determines the mass accretion rate. While hydrogen lines, such as H$\alpha$ at 656.3 nm, have been widely observed around young stars \citep[e.g.,][]{Joy1945TTS}, their flux relative to the continuum excess is significantly weaker---about two orders of magnitude lower in T Tauri stars \citep[e.g.,][]{Gullbring1998TTS}---and can often be negligible for estimating accretion rates.

When detecting continuum excess is challenging, especially for substellar objects with weaker excess, hydrogen lines are used to estimate the mass accretion rate \citep[e.g.,][]{Almendros-Abad2024YBD}. This is done through well-calibrated empirical relationships between accretion luminosity and hydrogen line emissions \citep{Herczeg2008ff,alca2017}. This method is preferred because the fluxes of these lines are concentrated in narrow wavelength ranges, making them easier to observe. 

\jhrev{Observations of substellar and planetary-mass companions have revealed unexpectedly high mass accretion rates in some cases \citep[e.g.,][]{zhou14,Betti2022D1b}. This can be attributed to the disk fragmentation formation scenario. Numerical simulations show that companions formed through disk fragmentation tend to have more massive circumsubstellar and circumplanetary disks, leading to higher accretion rates as they draw large amounts of gas from the massive protoplanetary disk of their parent star during the initial stages of their formation \citep{Stamatellos2015GI}.}

\jhrev{Furthermore,} for planetary-mass objects, the fraction of energy that escapes in hydrogen lines is much higher than for T Tauri stars \citep[e.g.,][]{zhou14,Zhou2021PDS70}, suggesting differences in accretion-emission mechanism. Two models have been developed to interpret the H$\alpha$ emission from these accreting planets: emission from the accretion flow \citep{Kwan2011Flow,Thanathibodee2019PDS70b}, which is thought to be the main origin of emission lines in accreting stars \citep{Hartmann2016accretion}, and emission from the shock \citep{Aoyama2018}. For the same H$\alpha$ luminosity, the estimated accretion rate can differ by an order of magnitude or more \citep[see Figure~2 in][]{Aoyama2021conversion}. For example, in the case of TWA~27~B with a mass of 5 $M_{\rm Jup}$ \citep{Chauvin2004}, the accretion rate estimated by the shock model is approximately 50 times higher than that estimated using extrapolated stellar scalings \citep{Marleau2024TWA27B}. Therefore, diagnostics are needed to determine which emission mechanism is dominant in individual planets.

One diagnostic for determining the emission mechanism is the line profile. For T Tauri stars, broad emission lines and/or red-shifted absorption components have been observed in some systems \citep[e.g.,][]{Bertout1982Accretion,Edwards1994Accretion,Muzerolle1998Accretion} and are interpreted as characteristic features in emission from the accreting inflows \citep[e.g.,][]{Calvet1992Model,Hartmann1994Model,Muzerolle1998Model,Wilson2022Model}. In contrast, the shock model predicts relatively symmetric line profiles \citep{Aoyama2018}. Hence, the red-shifted absorption feature, a.k.a., the inverse P-Cygni profile, could be a crucial diagnostic. However, if the accretion flows have sufficient temperature to absorb emission lines from shocks, the inverse P-Cygni profile could also be observed in emission lines from shocks. Additionally, the formation of the red-shifted absorption feature depends on radiative transfer and geometric effects \citep{Calvet1992Model,Hartmann1994Model}, meaning that the line profile may not always serve as a reliable diagnostic \citep{Edwards1994Accretion}.

Another approach to distinguish between the two models is by using line ratios. A pioneering study was conducted by \citet{Betti2022D1b}, who observed Delorme~1~(AB)b, a planet with a mass of 12-14 $M_{\rm Jup}$ \citep{Delorme2013D1b}, using multiple Paschen and Brackett lines obtained simultaneously. They found that the line ratios differ between emission from accretion flows and shocks, making them distinguishable. The observed line ratios of Delorme~1~(AB)b were compared to model predictions, suggesting that the emission mechanism of Delorme~1~(AB)b is best explained by the shock model. Subsequently, \citet{Aoyama2024TWA27B} applied this method to TWA 27 B, observed with JWST/NIRSPEC \citep{Luhman2023TWA27b}, although the uncertainties in the line ratios were too large to distinguish the emission mechanism.

For diagnosing the two models, simultaneous multi-line observations are preferable to obtain reliable line ratios free from variability \citep[e.g.,][]{Demars2023Variability}. However, the number of \jhrev{multi-line} observations for planetary-mass objects are very limited to date (e.g., SR 12 c; \citealp{Santamaria-Miranda2018SR12c}, Delorme 1 (AB)b; \citealp{Betti2022D1b}, TWA~27~B; \citealp{Aoyama2024TWA27B}). Thus, some questions remain unresolved. For example, is there a boundary between the models for planets, brown dwarfs, and stars? What physical parameters determine which model dominates?

In this paper, we apply two models to \jhrev{76} brown dwarfs and very low-mass stars with masses ranging from approximately 0.02 to 0.1 $M_\sun$ to address the questions mentioned above using archival data. As some objects were observed at multiple epochs, the total number of data points is 96. The structure of this paper is as follows: \S~\ref{sec:data} describes the archival data used in the paper, \S~\ref{sec:model} explains the flow and shock models, \S~\ref{sec:results} presents our results, and \S~\ref{sec:discuss} discusses relevant topics.

\section{Target Selection, Archive Data, and Observations} \label{sec:data}
\subsection{Target Selection and Archive data}\label{subsec:target}

Accreting brown dwarfs and very low-mass stars were selected from the CASPAR project\footnote{CASPAR is openly available on Zenodo:
\url{https://doi.org/10.5281/zenodo.8393054}.} \citep[the Comprehensive Archive of Substellar and Planetary Accretion Rates;][]{Betti2023Survey}. To study the accretion properties of these objects, we chose those with masses ranging from approximately 0.02 to 0.1 $M_\sun$. We excluded binary objects and those with an extinction of $A_V \gtrsim 3$~mag to avoid larger uncertainties in extinction correction.

Most of the multiple line fluxes for our selected targets have already been published in \citet{Herczeg2008ff,Alcala2014XSOOTER,alca2017,Venuti2019TWA} using Keck/LRIS \citep{Oke1995LRIS} and VLT/X-shooter \citep{Vernet2011Xshooter}. For other targets without published line fluxes, we retrieved calibrated 1D spectrum data from the SDSS\footnote{\url{https://www.sdss.org/}} \citep{York2000SDSS} and VLT/X-shooter\footnote{\url{http://archive.eso.org/eso/eso_archive_main.html}} archives. The wide wavelength coverage of these instruments allows for the simultaneous observation of multiple hydrogen lines. As higher excitation lines in the Paschen and Brackett series are generally too faint in lower-mass objects \citep[e.g.,][]{Natta2004YBD}, we focused on the Balmer lines in this paper. The spectral resolutions of the Balmer lines in the archival data are typically $R\sim2,000$ to 5,000.

As discussed in \S~\ref{sec:model}, we use H$\beta$ ($\lambda=4861.3$~\AA\ in air), H$\gamma$ ($\lambda=4340.5$~\AA), and H8 ($\lambda=3889.1$~\AA) to derive the line ratios. Therefore, we selected targets with an H8 line flux at a signal-to-noise ratio (SNR) greater than 3. Although H$\alpha$ ($\lambda=6562.8$ \AA) is the strongest line in the Balmer series, we excluded it because several factors beyond accretion---such as outflows, hot spots, chromospheric activity, complex magnetic field topology, and stellar rotation---can also affect its strength \citep[e.g.,][]{Alcala2014XSOOTER}. In addition to the CASPAR targets, we included objects from \citet{Theissen2017-2m1115,Theissen2018-2m1115,Almendros-Abad2024YBD} that meet these criteria. Table~\ref{tabA:physical} in \S~\ref{secA:properties} of the Appendix summarizes the physical properties of our \jhrev{76} targets.

Extinction in the line fluxes of H$\beta$, H$\gamma$, and H8 is corrected using the extinction law from \citet{Cardelli1989Extinction} with $R_{V} = 3.1$, except for the line fluxes listed in \citet{Alcala2014XSOOTER,alca2017, Venuti2019TWA}, which were already corrected by the authors. Since some objects were observed over multiple epochs, the total number of data points is 96, as summarized in Table~\ref{tabA:flux} in \S~\ref{secA:properties} of the Appendix.

\subsection{Subaru/HDS observations}\label{subsec:hds}

We carried out high-spectral-resolution observations with Subaru/HDS to investigate whether objects in different accretion categories, as described below, can be distinguished by their line profiles. On May 25, 2024 (UT), under program ID S24A-OT95 (PI: J. Hashimoto), we observed three objects---2MASS J15580252$-$3736026, 2MASS J16053215$-$1933159, and 2MASS J16083455$-$2211559---with total exposure times of 20, 60, and 60 minutes, respectively. The observations utilized the SETUP-Ba configuration, covering wavelengths from 3404.3 to 5081.5 \AA\ with a spectral resolution of $R = 36,000$. As discussed in \S~\ref{sec:results}, the emission lines for these objects are categorized as follows: flow-dominated for 2MASS J15580252$-$3736026, shock-dominated for 2MASS J16053215$-$1933159, and chromospheric-activity dominated for 2MASS J16083455$-$2211559. 

The data were reduced using the HDS reduction procedure `\verb#hdsql#', in conjunction with the Image Reduction and Analysis Facility (IRAF; \citealt{Tody1986IRAF}), as provided by the observatory\footnote{\url{https://www.naoj.org/Observing/Instruments/HDS/hdsql-e.html}}. The reduction process involved several steps: subtracting the average count from the over-scan region, correcting for scattered light and cross-talk, adjusting for detector nonlinearities, performing flat fielding, and removing cosmic rays.

Spectral extraction was performed using the \verb#apall# task in IRAF. The extracted spectra were flux-calibrated using the spectrophotometric standard star HR 7950\footnote{Data available here: \url{https://ftp.eso.org/pub/usg/standards/}}. Blaze-curvature correction was applied using nearby orders in the HR 7950 data, as the Balmer absorptions in HR 7950 are strong. Wavelength calibration was achieved with a Th-Ar comparison frame. Finally, the spectra were corrected for heliocentric radial velocity.

\section{Modeling Emission Lines} \label{sec:model}

We modeled the hydrogen line emission from both accretion-shock heated gas and the accretion flow. The shock emission model follows \citet{Aoyama2018}, which numerically simulates the post-shock cooling gas by solving 1D thermo-hydrodynamics, chemical reactions, and radiative transfer. The primary inputs for this model are the pre-shock flow velocity ($v_0$) and hydrogen-nuclei number density ($n_0$), with the output being the hydrogen line fluxes at the shock surface.

The accretion-flow emission is modeled following \citet{Kwan2011Flow}, treating the accretion flow as a slab with a constant temperature ($T$) and hydrogen-nuclei number density ($n_\mathrm{H}$). Using the Sobolev approximation, the optical depth and resulting escape probability are calculated based on the velocity-change length scale. For specific differences from \citet{Kwan2011Flow} and more details, please refer to Section 4.2.1 in \citet{Aoyama2024TWA27B}.

\begin{figure*}[htb!]
    \includegraphics[width=\linewidth]{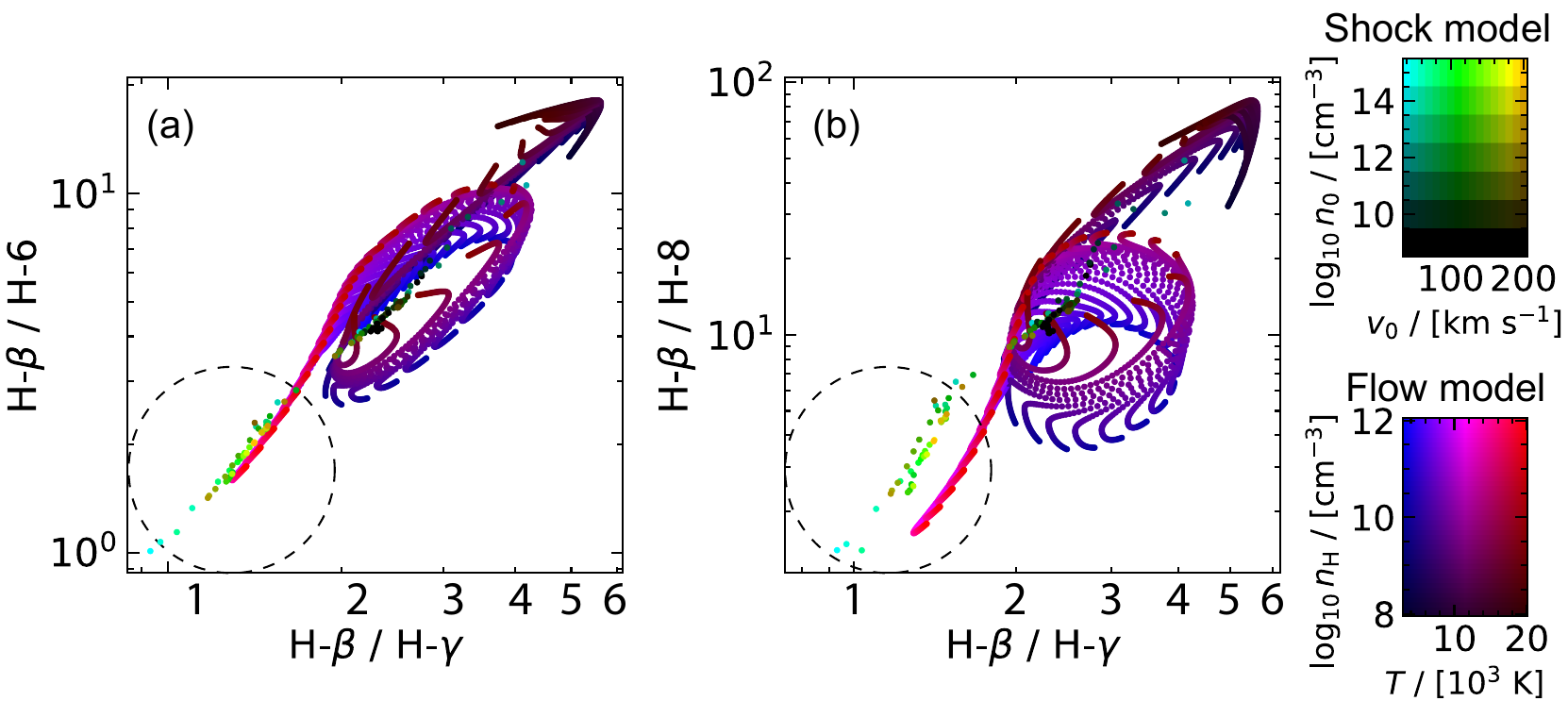} 
    \caption{
    The line ratio diagrams for (a) H$\beta$/H$\gamma$ versus H$\beta$/H6 and (b) H$\beta$/H$\gamma$ versus H$\beta$/H8. The loci of line ratios in the shock and flow models are deviated in the left bottom corner as highlighted with dashed lines, by including the H8 hydrogen line at higher upper level in panel (b). \jhrev{The loci of the shock model appear more discrete than those of the flow model because the parameter grid used in the shock model is coarser. However, this does not impact our overall results.} 
    }\label{fig:model}
\end{figure*}

Differences in hydrogen line emissions from shock-emission and accretion-flow mechanisms can be observed in the flux ratios among three or more hydrogen lines \citep{Betti2022D1b, Aoyama2024TWA27B}. While some ratios can be reproduced by both mechanisms with the appropriate choice of model input parameters, 
others are exclusive to one mechanism, reflecting the different behaviors in temperature and density changes.
%%%
The accretion flow is modeled as a slab because the temperature and density are expected to change only moderately during line emission. 
In contrast, the post-shock gas rapidly cools while emitting hydrogen lines, resulting in temperature and density variations over orders of magnitude. Consequently, the shock emission is a superposition of emissions from different temperatures and densities. Moreover, at a high $n_0$, radiative transfer, absorption, and re-emission in such rapidly varying gas cause the resulting line flux ratios to deviate further from those predicted by the slab model.

In this paper, we opt for a line-ratio diagram of H$\beta$/H$\gamma$ versus H$\beta$/H8 to differentiate the emission mechanisms. Figure~\ref{fig:model} shows that this combination of line ratios, which includes the higher excitation hydrogen line \jhrev{H8 ($\lambda=3889.1$ \AA)}, provides a greater distinction between shock and flow models compared to using the H6 line ($\lambda=4101.7$ \AA), especially in the lower left corner of the diagram. We exclude the H7 line ($\lambda=3970.1$ \AA) because it can be blended with the Ca II H line at $\lambda$3968 \AA. Additionally, higher excitation lines like H9 are not used due to their generally weaker signal, which results in lower signal-to-noise ratios (SNR).

The deviation observed in the H$\beta$/H$\gamma$ versus H$\beta$/H8 diagram (Figure \ref{fig:model}b) is due to the higher ionization degree in the flow model compared to the shock model. In both models, the populations of lower hydrogen levels, with principal quantum numbers of $n\lesssim6$, are primarily determined by collisional excitation, resulting in similar flux ratios for H$\beta$/H$\gamma$ and H$\beta$/H6. These ratios are primarily influenced by the temperature of the emitting regions. However, for higher hydrogen levels with $n\gtrsim7$, recombination of free electrons, in addition to collisional excitation, contributes to the level population. As a result, the H$\beta$/H8 ratio is lower in the flow model (Figure~\ref{fig:model}b).

\section{Results} \label{sec:results}

\subsection{Plot of line ratio}\label{subsec:plot}

\begin{figure}[htbp]
    \includegraphics[width=\linewidth]{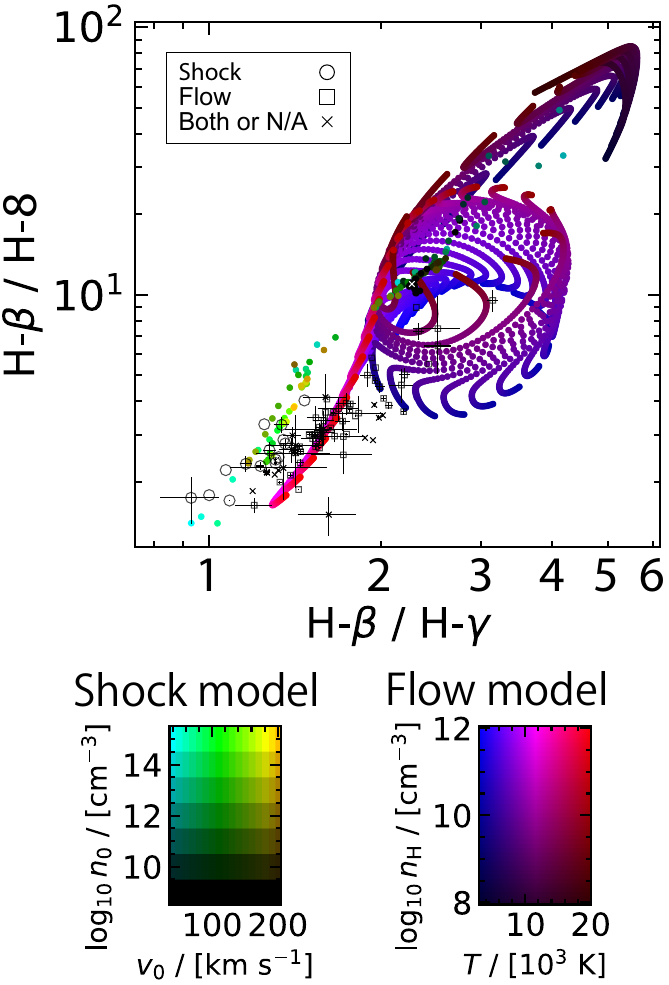} 
    \caption{
    Out of the 96 data points, 89 were analyzed for line ratios in the H$\beta$/H$\gamma$ versus H$\beta$/H8 diagram, with 7 points excluded due to dominant contributions from stellar chromospheric activity. These 89 data points are plotted against the loci of the shock and flow models (\S~\ref{sec:model}). The data are categorized as follows: 15 points are identified as shock-dominated emission (circles), 55 points as flow-dominated emission (squares), 13 points as fitting both models (crosses), and 6 points as not fitting either model (crosses).
    }\label{fig:plot-all}
\end{figure}

Figure~\ref{fig:plot-all} displays 89 data points of line ratios in the H$\beta$/H$\gamma$ versus H$\beta$/H8 diagram, overlaid on the two models described in \S~\ref{sec:model}. Seven out of 96 data points are excluded due to the potential dominance of stellar chromospheric activity in the hydrogen line flux, as discussed in \S~\ref{subsec:chr}. We classify the remaining 89 data points into four categories based on their 1$\sigma$ error bars: those that align with only the shock model, only the flow model, both models, or neither. For data points without reported errors or with errors less than 10\%, we assign a 10\% error to prevent classifying points near the model loci as `neither.' As a result, 15 data points are categorized as shock-dominated emission, 55 as flow-dominated emission, 13 as fitting both models, and 6 as not fitting either model, as summarized in Table~\ref{tabA:flux} in the Appendix.

\jhrev{The observed mix of emission mechanisms in brown dwarfs and very low-mass stars---15 and 55 points identified as shock- and flow-dominated emissions, respectively, as shown in Figure~\ref{fig:plot-all}---aligns with our expectations, as similar mixtures have been observed in T Tauri stars. \citet{Muzerolle1998Model} conducted numerical simulations of the accretion flow model, comparing the resulting H$\beta$/H$\gamma$ line ratios with observations of T Tauri stars. They found that some of these ratios (4 out of 16 T Tauri stars) could not be explained solely by the flow model and suggested that these emissions might originate from shock regions. While we cannot plot their line ratios in Figure~\ref{fig:plot-all} due to the absence of H8 line flux data in \citet{Muzerolle1998Model}, their lower H$\beta$/H$\gamma$ ratios align with the trend observed in Figure~\ref{fig:plot-all}, where shock-dominated emission is associated with reduced H$\beta$/H$\gamma$ ratios.} 

\jhrev{Assuming the emissions of these 4 T Tauri stars in \citet{Muzerolle1998Model} are attributed to shocks, Fisher's exact test shows that the difference in the fraction of shock- versus flow-dominated emissions between our 70 samples and the 16 T Tauri stars in \citet{Muzerolle1998Model} is statistically insignificant, at $\sim$0.3$\sigma$ with a two-tailed P-value of 74.6\%. Thus, our sample and the T Tauri stars are consistent.}

Fifteen out of the \jhrev{76} objects were observed at multiple epochs (see Table~\ref{tabA:flux} in the Appendix). Data points in 8 out of these 15 objects fall into different categories across different epochs. Notably, for five objects (2MASS J04322210$+$1827426, 2MASS J04390163$+$2336029, 2MASS J08440915$-$7833457, 2MASS J12073346$-$3932539, and 2MASS J16083081$-$3905488), the categorization alternates between the shock and flow models at different observational epochs.

\subsection{Line profile with Subaru/HDS}\label{subsec:hdsresults}

Although the line ratio is a useful diagnostic for determining the emission mechanism, the loci of the two models overlap at H$\beta$/H$\gamma \gtrsim2$, as shown in Figure~\ref{fig:plot-all}. As introduced in \S~\ref{sec:intro}, line profiles may help distinguish between the two models.

\begin{figure*}[htbp]
    \includegraphics[width=\linewidth]{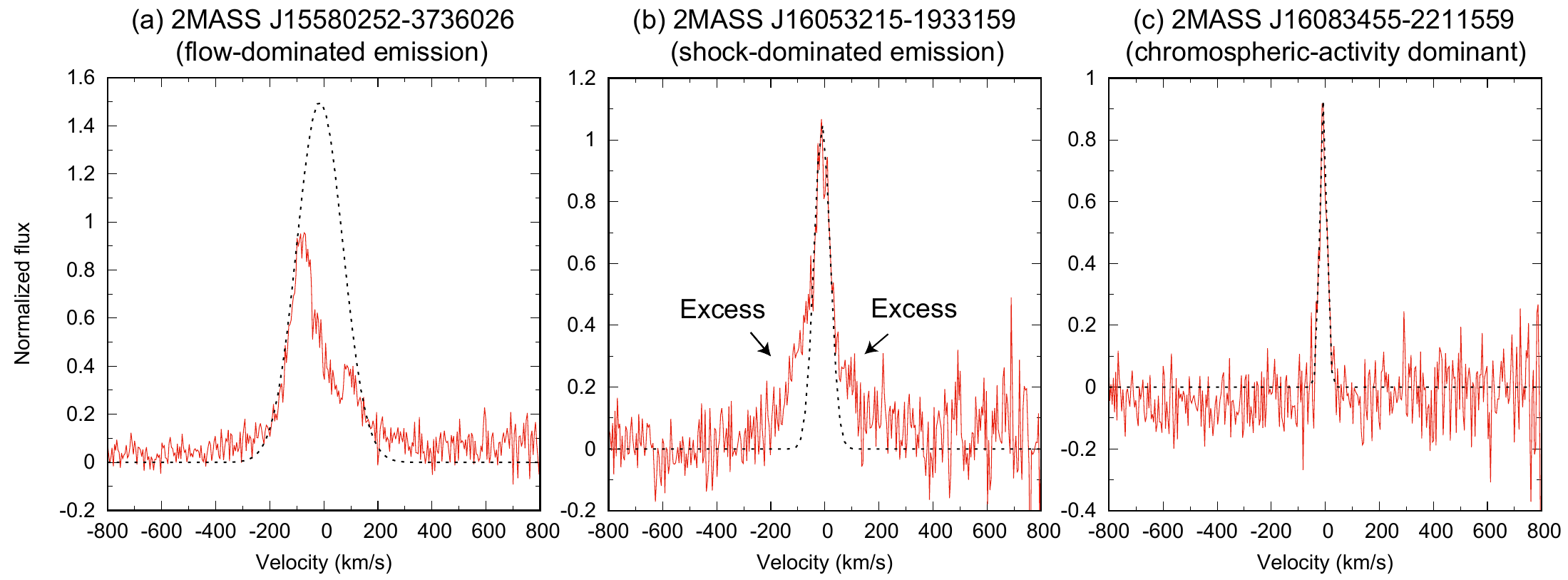} 
    \caption{
    The H$\beta$ spectra are shown for: (a) 2MASS J15580252$-$3736026 with flow-dominated emission, (b) 2MASS J16053215$-$1933159 with shock-dominated emission, and (c) 2MASS J16083455$-$2211559 with stellar chromospheric activity dominating. The black dotted lines represent Gaussian profiles with full widths at half maximum (FWHM) of 200.2 km/s for (a), 70.7 km/s for (b), and 30.6 km/s for (c), respectively.
    }\label{fig:hds}
\end{figure*}

Figure~\ref{fig:hds} shows the H$\beta$ spectra of three objects with a velocity resolution of 8.3~km/s: 2MASS J15580252$-$3736026 with flow-dominated emission, 2MASS J16053215$-$1933159 with shock-dominated emission, and 2MASS J16083455$-$2211559 with stellar chromospheric activity dominating. While H$\gamma$ and H$\delta$ emissions were also marginally detected, they are not shown due to lower SNR.

2MASS J15580252$-$3736026 (Figure~\ref{fig:hds}a) displays a line profile with blue- and red-shifted double peaks and an absorption feature at the line center. This profile is characteristic of when the accretion flow is oriented perpendicular to the line of sight \citep[e.g., Figure~4 in][]{Muzerolle1998Model}.

2MASS J16053215$-$1933159 (Figure~\ref{fig:hds}b) also shows an asymmetric line profile with a single peak at the line center, which may consist of multiple components. This type of profile has been observed in the planetary mass object Delorme 1 (AB)b, as explained by the shock model \citep{Betti2022D1b}. High spectral resolution observations with VLT/UVES at $R=50,000$ (6 km/s) revealed that the line profiles for Delorme 1 (AB)b included a combination of narrow and broad components with different velocity shifts \citep{Ringqvist2023Delorme1b}. The shock model generally predicts the single-peak profile. While high-resolution observations of shock-dominated emissions are limited to these examples, the presence of a single peak near the line center may consistently characterize shock-dominated emissions. However, a single peak profile can also be reproduced by the flow model in cases of high temperature \citep[e.g., Figure~4 in][]{Muzerolle2001Model}. Therefore, emission lines with a single peak profile can be attributed to accretion shocks and/or flows. In contrast, the presence of absorption features in the line profile suggests that the lines originate from accretion flows.

In contrast to the previous two cases, the line profile of 2MASS J16083455$-$2211559 displays a symmetric and narrow feature, with a Gaussian FWHM of 30.6~km/s (Figure~\ref{fig:hds}c). As discussed in \S~\ref{subsec:chr}, distinguishing between emission lines from accretion and those from chromospheric activity involves analyzing the accretion luminosity \citep{Manara2013Chr,Manara2017Chr}. Additionally, the equivalent width of H$\alpha$ is also used for this purpose \citep[e.g.,][]{White2003EW,Barrado2003EW}.

For weakly accreting objects, such as those at the end of their accretion phase, the above methods might categorize them as chromospheric-dominated. However, if these objects exhibit asymmetric line profiles such as Figures~\ref{fig:hds}(a) and (b), these could be evidence of accretion. Hence high spectral resolution observations could be particularly useful for identifying such weakly accreting objects. \jhrev{In addition to hydrogen lines, \citet{Thanathibodee2022WTTS} provide a catalog of weakly accreting objects, using \ion{He}{1} $\lambda$10830 as a sensitive probe of accretion. Intriguingly, the line ratio of 2MASS J16083455$-$2211559 falls within the shock model loci (\S~\ref{subsec:chr} and Figure \ref{figA:chr} of Appendix), suggesting it may be bona fide chromospheric-dominated.}

\section{Discussion} \label{sec:discuss}
\subsection{Chromospheric emission}\label{subsec:chr}

\begin{figure}[htbp]
    \includegraphics[width=\linewidth]{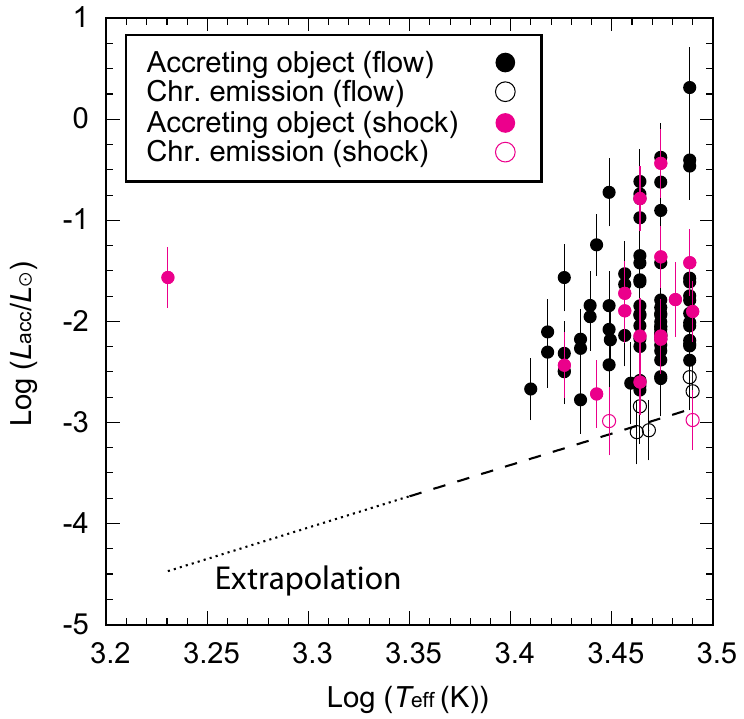} 
    \caption{
    The 96 data points for log($L_{\rm acc}/L_*$) are derived from the relationship between accretion luminosity and H$\beta$ line luminosity \citep{alca2017}, plotted against log($T_{\rm eff}{\rm /K}$). The empirical relationship for chromospheric activity in non-accreting stars is represented by a dashed line, validated within the range of $3.35$~dex~$\lesssim {\rm log} (T_{\rm eff}{\rm /K}) \lesssim 3.65$~dex, following the work of \citet{Manara2013Chr,Manara2017Chr}. For one object with log($T_{\rm eff}{\rm /K}) \approx 3.23$~dex, we extrapolate this relationship down to log($T_{\rm eff}{\rm /K}) = 3.2$~dex, indicated by the dotted line. Seven of these 96 data points are identified as dominated by chromospheric emission, as summarized in Table~\ref{tabA:flux} in the Appendix. 
    \jhrev{The 15 solid red circles represent data points categorized as shock-dominated emission in Figure \ref{fig:plot-all}, while the two open red circles indicate points identified as chromospheric emission that fall within the shock model loci in Figure \ref{figA:chr} of Appendix.}
    }\label{fig:chr}
\end{figure}

Chromospheric emissions can dominate line flux when mass accretion rates are low. \citet{Manara2013Chr,Manara2017Chr} explored the impact of chromospheric activity on mass accretion rate measurements. They developed an empirical relationship between the ratio of accretion luminosity to stellar luminosity ($L_{\rm acc}/L_*$) and the effective temperature ($T_{\rm eff}$) for non-accreting stars. This relationship serves as a diagnostic tool for assessing chromospheric activity, as illustrated in Figure~\ref{fig:chr}.

For our 96 data points, the accretion luminosity $L_{\rm acc}$ is calculated using a well-calibrated empirical conversion between accretion luminosity and H$\beta$ line luminosity \citep[Table B.1 in][]{alca2017}. Values for $L_*$ and $T_{\rm eff}$ are taken from the literature, as detailed in Table~\ref{tabA:physical}. We found that 7 of these 96 data points fall within 1$\sigma$ of the chromospheric activity locus shown in Figure~\ref{fig:chr}. Although the chromospheric activity relation for non-accreting stars is validated within $3.35$~dex~$\lesssim {\rm log} (T_{\rm eff}{\rm /K}) \lesssim$~3.65~dex \citep{Manara2013Chr,Manara2017Chr}, we extrapolate this relation to log($T_{\rm eff}{\rm /K}) = 3.2$~dex for the only coolest object (refer to \S~\ref{secA:2m1115} in Appendix). These seven points, marked in Table~\ref{tabA:flux} in the Appendix, will not be included in the subsequent analyses.

\jhrev{We note that \citet{Manara2013Chr,Manara2017Chr} used the $L_{\rm acc}$--$L_{\rm line}$ relation from \citet{alca2017} to derive the chromospheric activity relation. Therefore, in Figure \ref{fig:chr}, $L_{\rm acc}$ should be also derived using the relation from \citet{alca2017}. However, strictly speaking, the chromospheric activity relation in \citet{Manara2013Chr,Manara2017Chr} cannot be applied to objects with shock-dominated emission to determine whether they are dominated by chromospheric activity. That said, even if the $L_{\rm acc}$ of chromospheric activity is estimated using the $L_{\rm acc}$--$L_{\rm line}$ relation of the shock model \citep{Aoyama2021conversion}, the objects that fall on the chromospheric activity locus would remain the same as those with $L_{\rm acc}$ estimated using the relation from \citet{alca2017}, because the $L_{\rm line}$ used in estimating $L_{\rm acc}$ is same in both methods. In other words, the objects identified as chromospheric emission are consistent whether the relation from \citet{alca2017} or the shock model \citep{Aoyama2021conversion} is used.}

\begin{figure}[htbp]
    \includegraphics[width=\linewidth]{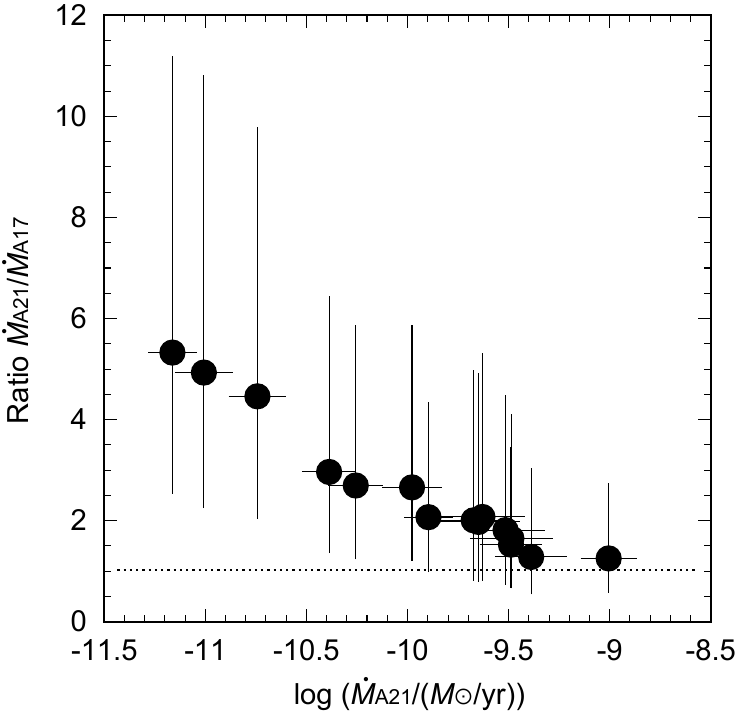} 
    \caption{
    Comparison of mass accretion rates derived using the shock model ($\dot{M}_{\rm A21}$) and the extrapolation of stellar scalings ($\dot{M}_{\rm A17}$) for 15 data points located on the shock model loci in Figure~\ref{fig:plot-all}. The smaller the mass accretion rate, the larger the deviation between the values of $\dot{M}_{\rm A21}$ and $\dot{M}_{\rm A17}$. Due to the 0.3-dex error in the conversion between $L_{H\beta}$ and $L_{\rm acc,A17}$ \citep{alca2017}, the error in the vertical axis is larger.
    }\label{fig:mdot}
\end{figure}

The line ratios of these 7 data points are plotted on the H$\beta$/H$\gamma$ versus H$\beta$/H8 diagram shown in Figure~\ref{figA:chr} in Appendix. \jhrev{Of these 7 points identified as chromospheric-dominated, two (2MASS J16083455$-$2211559 and 2MASS J11081850$-$7730408) fall within the shock model loci, and three (2MASS J04183030$+$2743208, 2MASS J04321786$+$2422149, and 2MASS J10561638$-$7630530) within the flow model loci (Figure \ref{figA:chr}, Appendix). These five objects may still be weakly accreting and could serve as valuable samples for understanding the final stages of accretion. Notably, 2MASS J16083455$-$2211559 was observed at a high spectral resolution of $R=36,000$ (8.3 km/s) with Subaru/HDS (\S~\ref{subsec:hds} and \S~\ref{subsec:hdsresults}).}

\subsection{Mass accretion rate}\label{subsec:mdot}

The estimates of accretion rates can vary significantly depending on whether the shock model or the extrapolation of stellar scalings, as introduced in \S~\ref{sec:intro}, is used. For the 15 data points located on the shock model loci in Figure~\ref{fig:plot-all}, their accretion rates should be estimated using the shock model. However, in the literature, these accretion rates are estimated using the Balmer continuum excess or the conversion between the Balmer continuum and hydrogen lines. In this subsection, we compare the accretion rate values derived from these two methods.

\begin{figure*}[htbp]
\centering
    \includegraphics[width=0.7\linewidth]{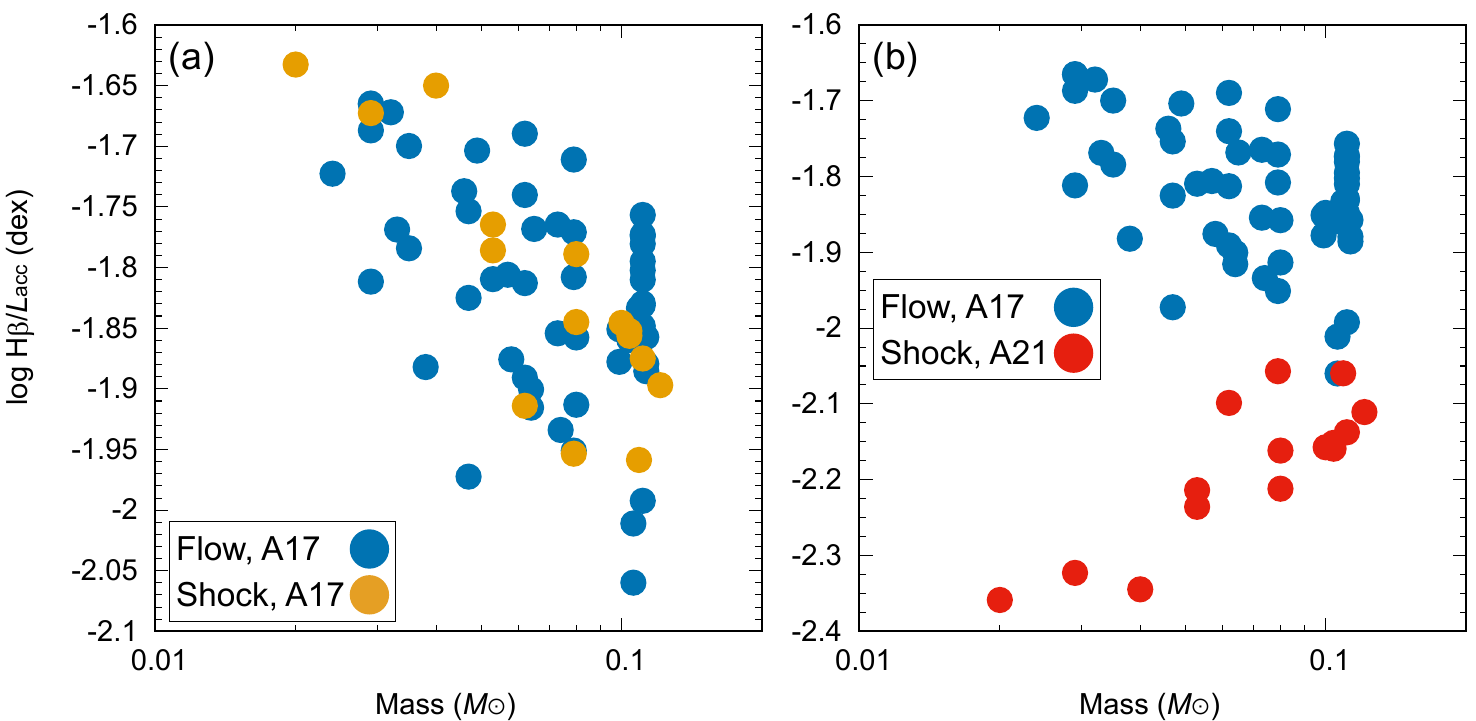} 
    \caption{
    The fraction of H$\beta$ line luminosity to accretion luminosity is shown as a function of mass for 70 data points, categorized as either flow-dominated (\textbf{blue}, 55 points) or shock-dominated (\textbf{yellow} in panel a and \textbf{red} in panel b, 15 points) emissions in Figure~\ref{fig:plot-all}. In panel (a), the accretion luminosity for the 15 shock-dominated data points is derived using stellar scalings \citep{alca2017}, while in panel (b), it is based on the shock model \citep{Aoyama2021conversion}. For clarity, the error bars are not shown in these panels; however, figures including the error are provided in Figure~\ref{figA:hb-lacc_error} in the Appendix.
    }\label{fig:hb_to_lacc}
\end{figure*}

The mass accretion rate $\dot{M}$ is estimated from the accretion luminosity $L_{\rm acc}$ using the following relation \citep{Gullbring1998TTS}:
\begin{equation}
\dot{M} \approx \frac{L_{\rm acc} R_*}{GM_*}, \label{eq:mdot}
\end{equation}
where $G$ is the gravitational constant, $M_*$ is the stellar mass, and $R_*$ is the stellar radius (as summarized in Table~\ref{tabA:physical} in Appendix). The values derived from the shock model in \citet{Aoyama2021conversion} are denoted by $\dot{M}_{\rm A21}$ and $L_{\rm acc,A21}$, while those derived from stellar scalings in \citet{alca2017} are denoted by $\dot{M}_{\rm A17}$ and $L_{\rm acc,A17}$. In both cases, the H$\beta$ line luminosity $L_{H\beta}$ is converted to the accretion luminosity.

Figure~\ref{fig:mdot} compares the two accretion rates, $\dot{M}_{\rm A21}$ and $\dot{M}_{\rm A17}$. The values are similar when the accretion rate is greater than approximately 10$^{-9.5}M_\sun$/yr. Meanwhile, for smaller mass accretion rates, the rate derived using the shock model is systematically higher than that estimated using stellar scalings. Generally, the mass accretion rate is proportional to the mass, following the empirical relation $\dot{M} \propto M^2$ in both the stellar and sub-stellar regimes \citep[e.g.,][]{Betti2023Survey}. Assuming this relation extends to the planetary mass regime ($M \lesssim 10 M_{\rm Jup}$) without a change in slope, deviations in the mass accretion rates of planetary mass objects could be larger. Therefore, it is crucial to determine which emission mechanism is dominant, especially for sub-stellar objects with smaller mass accretion rates.

\jhrev{For reference, in addition to the mass accretion rate, we also compare the accretion luminosity derived using two methods, $L_{\rm acc,A21}$ and $L_{\rm acc,A17}$, in Figure \ref{figA:Lacc} of Appendix \ref{secA:Lacc}. Overall, the general trend of $L_{\rm acc,A21}$/$L_{\rm acc,A17}$ is consistent with $\dot{M}_{\rm A21}$/$\dot{M}_{\rm A17}$.}

\subsection{Fraction of H$\beta$ luminosity to accretion luminosity}\label{subsec:fraction}

The fraction of energy emitted in hydrogen lines is reported to be significantly higher for objects with lower accretion luminosity \citep[e.g.,][]{zhou14,Zhou2021PDS70}. This relationship also applies to mass, as accretion luminosity is proportional to mass \citep[e.g., Figure 15 in][]{Herczeg2008ff}. Figure~\ref{fig:hb_to_lacc}(a) illustrates the fraction of H$\beta$ line luminosity relative to accretion luminosity for 70 data points, categorized as either flow-dominated or shock-dominated emissions (Figure~\ref{fig:plot-all}), plotted as a function of mass. The accretion luminosity for all data points is derived from stellar scaling relations \citep{alca2017}. As expected, Figure~\ref{fig:hb_to_lacc}(a) confirms this relationship. For clarity, error bars are omitted from the main figure but can be found in Figure~\ref{figA:hb-lacc_error} in the Appendix.

In Figure~\ref{fig:hb_to_lacc}(b), the accretion luminosity for 15 data points, classified as shock-dominated emissions, is estimated using the shock model \citep{Aoyama2021conversion}. A clear bimodal distribution emerges when comparing accretion luminosities derived from stellar scaling relations and those from the shock model. The fraction of H$\beta$ line luminosity to accretion luminosity, as calculated by the shock model, is systematically lower than that derived from stellar scaling. The deviation between these two approaches becomes more pronounced as the mass decreases. This is expected, as the fraction is described as
\begin{equation}
{\rm log}~L_{\rm H\beta}/L_{\rm acc}~\propto~\frac{1-a}{a}~{\rm log}~L_{\rm acc},  \label{eq:hb-lacc}
\end{equation}
where $(1-a)/a$ is positive for the shock model \citep[see Table~1 in][]{Aoyama2021conversion} and negative for stellar scalings \citep[see Table B.1 in][]{alca2017}. This trend may apply to all hydrogen lines. These findings suggest that the efficiency of line emission is higher in accretion flows compared to the shock region, particularly in the substellar regime.

\begin{figure*}[htbp]
    \includegraphics[width=\linewidth]{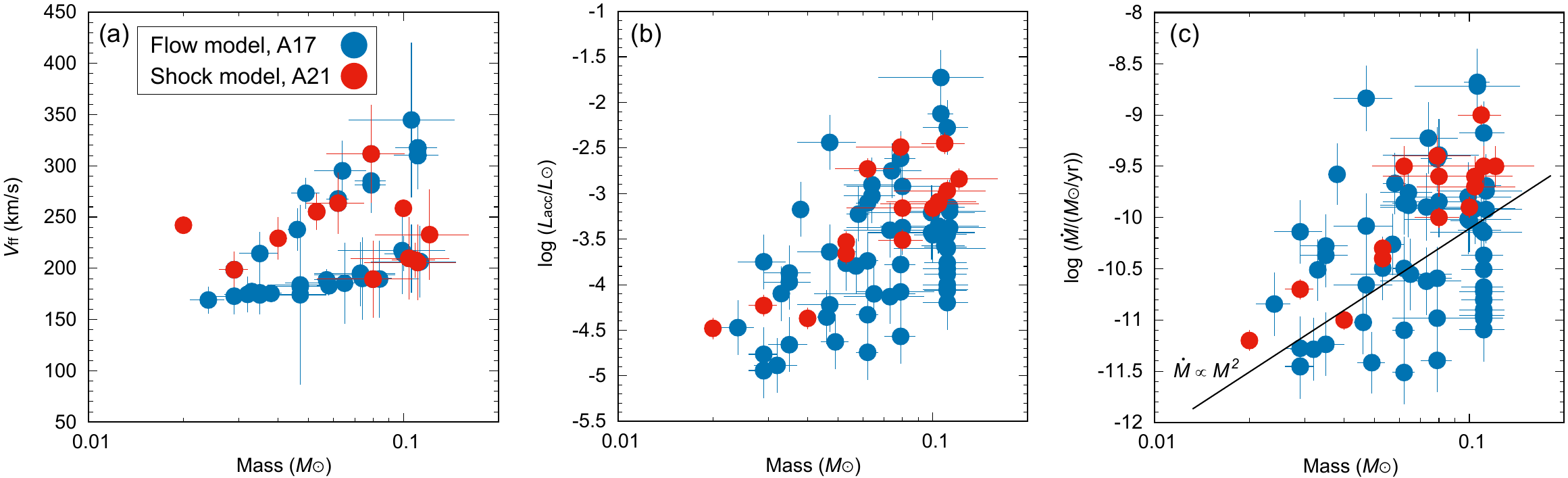} 
    \caption{
    The 70 data points, categorized as either flow-dominated (blue, 55 points) or shock-dominated (red, 15 points) emissions in Figure~\ref{fig:plot-all}, are superimposed on diagrams depicting accretion-related physical quantities: (a) free-fall velocity ($v_{\rm ff}$) from infinity, (b) accretion luminosity ($L_{\rm acc}$), and (c) accretion rate ($\dot{M}$), all plotted as functions of mass. A black solid line represents the empirical relation $\dot{M} \propto M^2$ across both stellar and substellar regimes \citep[e.g.,][]{Betti2023Survey}. There appears to be no clear boundary between the shock and flow model categories.
    }\label{fig:param}
\end{figure*}

\subsection{Boundary between the flow and shock models}\label{subsec:boundary}

One of the purposes of this paper is to identify the physical quantity that determines the boundary between the flow and shock models. Figure~\ref{fig:param} presents 70 data points: 15 categorized under the shock model and 55 under the flow model, as shown in Figure~\ref{fig:plot-all}. These points are plotted on diagrams illustrating accretion-related physical quantities, including free-fall velocity ($v_{\rm ff}$) from infinity, accretion luminosity ($L_{\rm acc}$), and accretion rate ($\dot{M}$), all as functions of mass. The values of $L_{\rm acc}$ and $\dot{M}$ are estimated as outlined in \S~\ref{subsec:mdot}. As shown in Figure~\ref{fig:param}, there appears to be no clear boundary between the shock and flow model categories.

Regarding the free-fall velocity from infinity in Figure~\ref{fig:param}(a), we anticipate that higher free-fall velocities correspond to greater shock energy and increased viscous heating in accretion flows. This should result in brighter emission lines in both accretion flows and shock regions. However, since no clear boundary is observed relative to $v_{\rm ff}$, it does not appear to selectively enhance the brightness of emission lines from flows and shocks. Additionally, as accretion flows are launched from the inner truncated region of the disk, the effective free-fall velocity depends on the truncation radius of the disks. Although the truncation radius is often assumed to be around 5~$R_*$ \citep[e.g.,][]{Hartmann2016accretion}, it may vary between objects. Therefore, it may be necessary to consider the practical truncation radius when determining the boundary.

\begin{figure*}[htbp]
    \includegraphics[width=\linewidth]{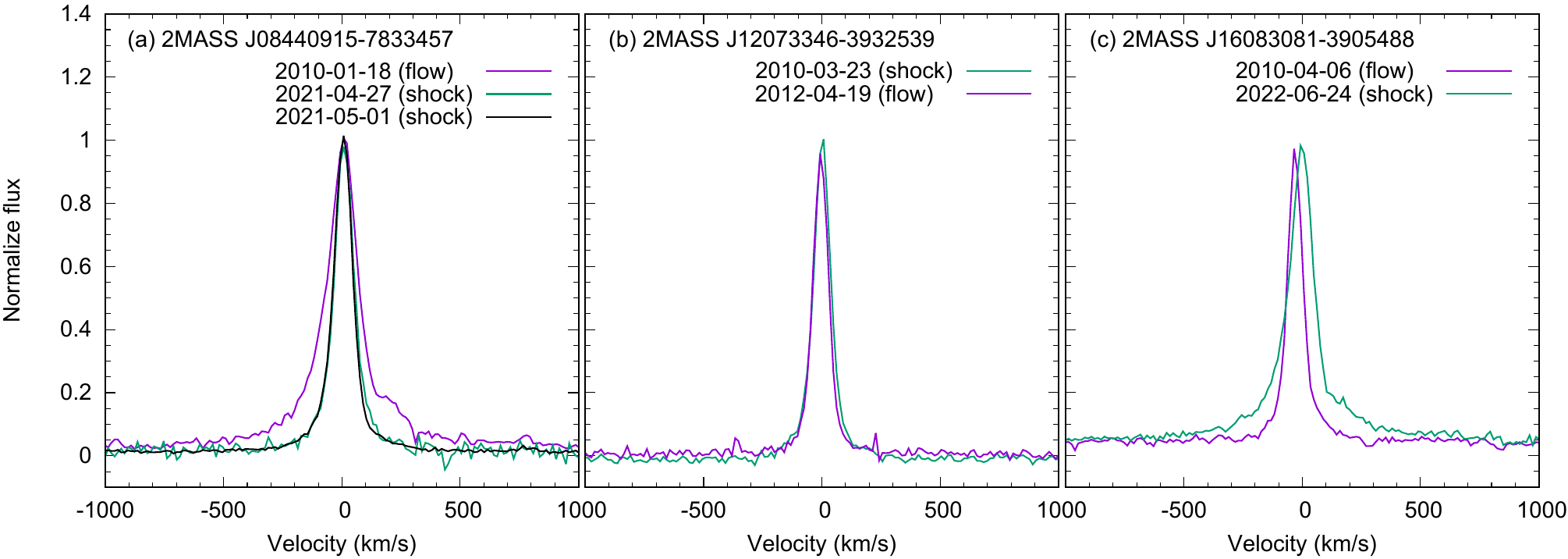} 
    \caption{
    Comparison of H$\gamma$ line profiles for objects observed at multiple epochs, where the emission categories switched between shock and flow models. The spectra were obtained using VLT/Xshooter with a resolution of $R\approx5,400$ (55.6 km/s).
    }\label{fig:profile}
\end{figure*}

For the accretion luminosity in Figure~\ref{fig:param}(b), the fraction of energy emitted as hydrogen lines is significantly higher in objects with lower accretion luminosity \citep[e.g.,][]{zhou14,Zhou2021PDS70}, possibly reflecting differences in emission mechanisms. Indeed, the fraction of H$\beta$ line luminosity relative to accretion luminosity is systematically higher for flow-dominated emissions compared to shock-dominated emissions among very low-mass stars and brown dwarfs in our sample (Figure~\ref{fig:hb_to_lacc}b). However, no distinct boundary between the two categories is observed in Figure~\ref{fig:param}(b).

Similarly to accretion luminosity, we anticipated the presence of boundary for the mass accretion rate (Figure~\ref{fig:param}c) since the accretion rate is proportional to the accretion luminosity (Eq.~\ref{eq:mdot}). The factor of $R_*/M_*$ in Eq.~\ref{eq:mdot} varies by a factor of less than 10, the general trend between the accretion rate and accretion luminosity remains consistent, as shown in Figure~\ref{figA:difference} in Appendix. Overall, it remains unclear which of the four parameters is the primary determinant of the emission mechanism.

\subsection{Time variability of line ratios}\label{subsec:boundary}

Five objects were observed across multiple epochs, with their positions on the shock and flow models varying between epochs (see Table~\ref{tabA:flux} in the Appendix). For H$\beta$ line luminosity, no clear trend emerges indicating which mechanism---shock or flow---consistently produces brighter emissions. In two of the five objects, the accretion flow resulted in brighter line emissions, while in the other three, the shock mechanism dominated. This variability may be due to fluctuations in the emission lines, a phenomenon that has been documented in both stellar and substellar regimes \citep[e.g.,][]{Fischer2023PPVII_variability,Demars2023Variability}.

We also examine the line profiles because emission lines from accreting flows are expected to show a red-shifted absorption component around 100-300 km/s \citep[e.g.,][]{Calvet1992Model} and/or an absorption feature at the line center \citep[e.g.,][]{Muzerolle1998Model}, whereas the shock model predicts symmetric emission lines \citep{Aoyama2018}. Calibrated spectra for three of the five objects (2MASS J08440915$-$7833457, 2MASS J12073346$-$3932539, and 2MASS J16083081$-$3905488), obtained with VLT/Xshooter at $R\approx5,400$ (55.6 km/s), are publicly available on the ESO archival website. Figure~\ref{fig:profile} displays the H$\gamma$ line profiles of these three objects at different epochs. None of the spectra categorized as flow models clearly exhibit the red-shifted absorption component or the absorption feature at the line center. Given that the velocity resolution of 55.6 km/s should be sufficient to resolve a red-shifted absorption component at approximately 100-300 km/s, the absence of this feature could be attributed to the complexities of radiative transfer and geometric effects, which influence the formation of the red-shifted absorption feature \citep{Calvet1992Model,Hartmann1994Model}. Additionally, if the absorption feature is both narrow and shallow, it may not be detectable with the current spectral resolution. Similarly, the absorption feature at the line center might also remain unresolved given the limitations of the current spectral resolution.

Changes in line width between the two models (Figure~\ref{fig:profile}) could potentially serve as a diagnostic tool. An unresolved absorption feature at the line center might result in a broader line width. However, this hypothesis may not hold up in practice. For example, the flow model exhibits a wider line width for 2MASS J08440915$-$7833457, while the shock model shows a wider line width for 2MASS J16083081$-$3905488. In the case of 2MASS J12073346$-$3932539, the line widths are similar for both models. Therefore, line ratios may provide a more reliable means of distinguishing between the two models.

The reason for the switch in emission mechanisms remains unclear, but it could be related to temperature changes in the accretion flow. If the temperature in accretion flows decreases---potentially due to reduced viscous heating---shock emissions might become relatively brighter compared to flow emissions. \jhrev{Alternatively, other mechanisms, such as extinction by accretion flows and flare activity, may influence line variability. However, distinguishing the origin of this variability is not always straightforward. Possible mechanisms are summarized in Table 1 and Figure 3 of \citet{Fischer2023PPVII_variability}.}

\section{Summary}

The mass accretion rate can be estimated from hydrogen emission lines using two distinct models: the accretion flow model and the accretion shock model. The accretion flow model assumes that hydrogen emissions primarily come from the accretion flow, whereas the accretion shock model attributes emissions to the shock region. As a result, these models can yield different estimates for the mass accretion rate. The discrepancy between the estimates tends to increase with decreasing mass accretion rate. Therefore, it is crucial to determine which emission mechanism is dominant, particularly for planetary mass objects. To distinguish between the two models, line ratios can be used as diagnostic tools. In this context, we applied both models to archival data from 96 data points representing \jhrev{76} accreting brown dwarfs and very low-mass stars, with masses ranging from approximately 0.02 to 0.1 $M_\sun$, to assess which model better explains their accreting properties. Our findings are summarized below:

\begin{itemize}
  
  \item Out of the 96 data points, 89 were analyzed after excluding 7 points due to the dominant contributions of stellar chromospheric activity. These 89 data points were plotted on the loci of the shock and flow models in the H$\beta$/H$\gamma$ versus H$\beta$/H8 diagram. The data points were categorized as follows: 15 as shock-dominated emissions, 55 as flow-dominated emissions, 13 as exhibiting both types of emissions, and 6 as neither. 

  \item For the 15 data points explained by the planetary shock model, we compared the mass accretion rates estimated by the shock model and the extrapolation of stellar scalings. We found that the shock model estimates mass accretion rates that are up to several times higher than those estimated by the stellar scalings. This difference, which can range from one to two orders of magnitude for planetary mass objects, underscores the importance of determining the dominant emission mechanism for each individual planet.

  \item The fraction of line luminosity to accretion luminosity increases as mass decreases when using stellar scalings, while the opposite trend is observed in the shock model. This indicates that the efficiency of line emissions is higher in accretion flows compared to the shock region within the substellar regime.

  \item We aimed to identify boundaries between the two models based on accretion-related physical quantities, including object mass, free-fall velocity from infinity, accretion luminosity, and accretion rate. However, no distinct boundary between the objects categorized under the two models was found.
  
  \item High-resolution spectroscopy with a resolution greater than $R > 10,000$ could help distinguish between line profiles dominated by flow and those dominated by shock emissions. Flow-dominated emissions may exhibit absorption components, whereas shock-dominated emissions typically present a single peak at the line center. However, more observations are needed, as only two objects with shock-dominated emissions have been observed with high-resolution spectroscopy to date.
  
  \item We observed variability in line ratios across different epochs for five objects, which were plotted in different loci of the shock and flow models at different epochs. Although we anticipated differences in line profiles due to changes in emission mechanisms, the profiles generally remained similar. The reason for the switch between emission mechanisms remains unclear.
  
\end{itemize}

\jhrev{Finally, we briefly discuss future prospects for direct imaging observations of accreting planets. Our studies demonstrate that the emission mechanisms in brown dwarfs and very low-mass stars can be described by either flow models, originally developed for accreting stars, or shock models, initially created for accreting planets, with the flow model being more prevalent. This finding suggests that accreting planets can also be explained using both models. Depending on the emission mechanism, line flux can vary even with the same accretion luminosity (or accretion rate). Although the efficiency of line emission is higher in accretion flows, particularly for planetary mass objects with low accretion luminosity, the observed emission line from the flow is not always brighter than that from the shock. This indicates that the detection rate of emission lines from accreting planets is independent of the emission mechanism.}

\newpage

\begin{acknowledgments}

\jhrev{The authors thank the anonymous referee for a timely and constructive report.}
This study was supported by JSPS KAKENHI Grant Number 23K03463.
This research is based [in part] on data collected at the Subaru Telescope, which is operated by the National Astronomical Observatory of Japan. We are honored and grateful for the opportunity of observing the Universe from Maunakea, which has the cultural, historical and natural significance in Hawaii. 
Based on observations collected at the European Southern Observatory under ESO programme(s) 084.C-0269(B), 084.C-1095(A), 085.C-0238(A), 085.C-0876(A), 086.C-0173(A), 087.C-0244(A), 089.C-0143(A), 093.C-0097(A), 093.C-0109(A), 093.C-0769(A), 094.C-0805(A), 095.C-0134(A), 095.C-0378(A), 096.C-0455(A), 0101.C-0527(A), 0101.C-0866(A), 0104.C-0454(A), 106.20Z8.008, 108.22CB.001, and/or data obtained from the ESO Science Archive Facility with DOI(s) under https://doi.org/10.18727/archive/71.
Funding for the Sloan Digital Sky Survey (SDSS) has been provided by the Alfred P. Sloan Foundation, the Participating Institutions, the National Aeronautics and Space Administration, the National Science Foundation, the U.S. Department of Energy, the Japanese Monbukagakusho, and the Max Planck Society. The SDSS Web site is http://www.sdss.org/.
The SDSS is managed by the Astrophysical Research Consortium (ARC) for the Participating Institutions. The Participating Institutions are The University of Chicago, Fermilab, the Institute for Advanced Study, the Japan Participation Group, The Johns Hopkins University, Los Alamos National Laboratory, the Max-Planck-Institute for Astronomy (MPIA), the Max-Planck-Institute for Astrophysics (MPA), New Mexico State University, University of Pittsburgh, Princeton University, the United States Naval Observatory, and the University of Washington.
IRAF is distributed by the National Optical Astronomy Observatories, which are operated by the Association of Universities for Research in Astronomy, Inc., under cooperative agreement with the National Science Foundation.

\end{acknowledgments}

\software{
          astropy \citep{Astropy2013,Astropy2018,Astropy2022},  
          IRAF \citep{Tody1986IRAF},
          Matplotlib \citep{Hunter2007Matplotlib},
          Numpy \citep{VanDerWalt2011}
          }

\facilities{Keck, VLT, SDSS, Subaru}

\appendix

\section{Physical and accretion properties of targets}\label{secA:properties}

Physical and accretion properties of objects are summarized in Table~\ref{tabA:physical} and \ref{tabA:flux}. 

%\clearpage

\begin{startlongtable}
\begin{deluxetable*}{ccccccccl}
\tablewidth{0pt} 
%\tablenum{1}
\tablecaption{Physical properties of targets.} \label{tabA:physical}
\tablehead{
\colhead{2MASS Name} &\colhead{$M_*$}     &\colhead{$R_*$}     &\colhead{log ($L_*/L_\sun$)} &\colhead{log ($T_{\rm eff}$/K)} &\colhead{$v_{\rm ff}$} & \colhead{$A_V$} &\colhead{$d$} &\colhead{Refs} \\
\colhead{}           &\colhead{($M_\sun$)}&\colhead{($R_\sun$)}&\colhead{(dex)}          &\colhead{(dex)}                   &\colhead{(km/s)} &\colhead{(mag)}&\colhead{(pc)}&\colhead{} 
}
\decimalcolnumbers
\startdata
J04141188$+$2811535             & 0.047 $\pm$ 0.010 & 0.59 $\pm$ 0.08 & -1.717 $\pm$ 0.153 & 3.449 & 174.3 $\pm$ 87.9 & 1.0 & 130.4 & 1,1,1,1,1\\
J04183030$+$2743208             & 0.084 $\pm$ 0.028 & 0.89 $\pm$ 0.19 & -1.290 $\pm$ 0.217 & 3.464 & 189.7 $\pm$ 37.5 & 0.3 & 128.0 & 1,1,1,1,1\\
J04262939$+$2624137             & 0.057 $\pm$ 0.005 & 0.61 $\pm$ 0.04 & -1.655 $\pm$ 0.067 & 3.456 & 188.7 $\pm$ 10.3 & 0.8 & 157.3 & 1,1,1,1,1\\
J04321606$+$1812464             & 0.058 $\pm$ 0.005 & 0.66 $\pm$ 0.04 & -1.593 $\pm$ 0.068 & 3.456 & 183.0 $\pm$  9.6 & 0.3 & 148.5 & 1,1,1,1,1\\
J04321786$+$2422149$^a$         & 0.07              & 0.61            & -1.602             & 3.468 & 209.1            & 0.0 & 126.5 & 2,2,2,2,2\\
J04322210$+$1827426             & 0.109 $\pm$ 0.017 & 0.96 $\pm$ 0.05 & -1.188 $\pm$ 0.075 & 3.474 & 208.0 $\pm$ 17.1 & 1.0 & 145.6 & 1,1,1,1,1\\
J04362151$+$2351165             & 0.104 $\pm$ 0.033 & 0.90 $\pm$ 0.19 & -1.240 $\pm$ 0.217 & 3.474 & 209.9 $\pm$ 40.0 & 1.0 & 114.5 & 1,1,1,1,1\\
J04382134$+$2609137             & 0.073 $\pm$ 0.022 & 0.73 $\pm$ 0.11 & -1.460 $\pm$ 0.162 & 3.464 & 195.2 $\pm$ 32.9 & 2.0 & 139.4 & 1,1,1,1,1\\
J04390163$+$2336029             & 0.111 $\pm$ 0.036 & 1.00 $\pm$ 0.15 & -1.813 $\pm$ 0.139 & 3.489 & 205.7 $\pm$ 36.8 & 0.0 & 126.8 & 1,1,1,1,1\\
J04390396$+$2544264             & 0.035 $\pm$ 0.006 & 0.44 $\pm$ 0.06 & -2.017 $\pm$ 0.155 & 3.439 & 174.1 $\pm$ 19.1 & 0.8 & 140.9 & 1,1,1,1,1\\
J04414825$+$2534304             & 0.029 $\pm$ 0.005 & 0.37 $\pm$ 0.04 & -2.183 $\pm$ 0.147 & 3.427 & 172.8 $\pm$ 17.6 & 1.0 & 137.3 & 1,1,1,1,1\\
J04432023$+$2940060             & 0.100 $\pm$ 0.013 & 0.83 $\pm$ 0.08 & -1.319 $\pm$ 0.164 & 3.474 & 214.3 $\pm$ 17.3 & 0.3 & 156.0 & 1,1,1,1,1\\
J04442713$+$2512164             & 0.035 $\pm$ 0.006 & 0.43 $\pm$ 0.06 & -2.030 $\pm$ 0.154 & 3.439 & 176.1 $\pm$ 19.5 & 0.0 & 141.9 & 1,1,1,1,1\\
J04554757$+$3028077             & 0.121 $\pm$ 0.040 & 0.85 $\pm$ 0.16 & -1.269 $\pm$ 0.214 & 3.481 & 232.9 $\pm$ 44.3 & 1.0 & 156.0 & 1,1,1,1,1\\
J05180285$+$2327127             & 0.062 $\pm$ 0.013 & 0.34 $\pm$ 0.03 & -2.129 $\pm$ 0.120 & 3.464 & 263.6 $\pm$ 30.0 & 1.0 & 103.8 & 1,1,1,1,1\\
J05382358$-$0220475             & 0.074 $\pm$ 0.018 & 0.78 $\pm$ 0.27 & -1.403 $\pm$ 0.400 & 3.464 & 190.2 $\pm$ 40.2 & 0.5 & 410.0 & 1,1,1,1,1\\
J05382543$-$0242412             & 0.038 $\pm$ 0.002 & 0.47 $\pm$ 0.02 & -1.932 $\pm$ 0.052 & 3.442 & 175.6 $\pm$  5.9 & 0.7 & 409.2 & 1,1,1,1,1\\
J05384818$-$0244007             & 0.106 $\pm$ 0.010 & 0.92 $\pm$ 0.28 & -1.222 $\pm$ 0.293 & 3.474 & 209.6 $\pm$ 33.4 & 0.8 & 433.3 & 1,1,1,1,1\\
J05395173$-$0222472             & 0.106 $\pm$ 0.010 & 0.92 $\pm$ 0.28 & -1.222 $\pm$ 0.293 & 3.474 & 209.6 $\pm$ 33.4 & 1.5 & 398.4 & 1,1,1,1,1\\
J08361073$-$7908184             & 0.062 $\pm$ 0.005 & 0.33 $\pm$ 0.04 & -2.158 $\pm$ 0.109 & 3.464 & 267.6 $\pm$ 19.5 & 0   &  94.0 & 1,1,1,1,1\\
J08385150$-$7916137             & 0.062 $\pm$ 0.005 & 0.33 $\pm$ 0.04 & -2.158 $\pm$ 0.109 & 3.464 & 267.6 $\pm$ 19.5 & 0   &  94.0 & 1,1,1,1,1\\
J08413030$-$7853064             & 0.079 $\pm$ 0.009 & 0.37 $\pm$ 0.04 & -2.021 $\pm$ 0.110 & 3.474 & 285.3 $\pm$ 22.4 & 0   &  98.9 & 1,1,1,1,1\\
J08422710$-$7857479             & 0.111 $\pm$ 0.017 & 0.42 $\pm$ 0.05 & -1.841 $\pm$ 0.115 & 3.489 & 317.4 $\pm$ 30.8 & 0   &  98.5 & 1,1,1,1,1\\
J08440915$-$7833457             & 0.053 $\pm$ 0.003 & 0.31 $\pm$ 0.04 & -2.236 $\pm$ 0.110 & 3.456 & 255.3 $\pm$ 18.0 & 0   &  99.6 & 1,1,1,1,1\\
J08441637$-$7859080             & 0.111 $\pm$ 0.017 & 0.42 $\pm$ 0.05 & -1.841 $\pm$ 0.115 & 3.489 & 317.4 $\pm$ 30.8 & 0   &  94.0 & 1,1,1,1,1\\
J10561638$-$7630530             & 0.076 $\pm$ 0.008 & 0.84 $\pm$ 0.12 & -1.350 $\pm$ 0.115 & 3.462 & 185.7 $\pm$ 16.5 & 0   & 190.9 & 1,1,1,1,1\\
J11020983$-$3430355             & 0.023 $\pm$ 0.002 & 0.26 $\pm$ 0.02 & -2.595 $\pm$ 0.068 & 3.410 & 183.6 $\pm$ 10.7 & 0   &  59.2 & 1,1,1,1,1\\
J11064180$-$7635489             & 0.113 $\pm$ 0.013 & 1.01 $\pm$ 0.11 & -1.142 $\pm$ 0.100 & 3.474 & 206.5 $\pm$ 16.4 & 0   & 194.7 & 1,1,1,1,1\\
J11065939$-$7530559             & 0.080 $\pm$ 0.027 & 0.85 $\pm$ 0.18 & -1.332 $\pm$ 0.226 & 3.464 & 189.4 $\pm$ 37.7 & 0.4 & 192.5 & 1,1,1,1,1\\
J11081850$-$7730408             & 0.047 $\pm$ 0.010 & 0.58 $\pm$ 0.09 & -1.732 $\pm$ 0.157 & 3.449 & 175.8 $\pm$ 23.1 & 0   & 186.7 & 1,1,1,1,1\\
J11082238$-$7730277             & 0.080 $\pm$ 0.013 & 0.85 $\pm$ 0.10 & -1.332 $\pm$ 0.118 & 3.464 & 189.4 $\pm$ 19.0 & 1.3 & 190.1 & 1,1,1,1,1\\
J11083952$-$7734166             & 0.065 $\pm$ 0.023 & 0.72 $\pm$ 0.17 & -1.491 $\pm$ 0.277 & 3.459 & 185.5 $\pm$ 39.5 & 0.3 & 187.4 & 1,1,1,1,1\\
J11085090$-$7625135             & 0.080 $\pm$ 0.027 & 0.85 $\pm$ 0.18 & -1.332 $\pm$ 0.226 & 3.464 & 189.4 $\pm$ 37.7 & 0.8 & 192.2 & 1,1,1,1,1\\
J11120984$-$7634366             & 0.113 $\pm$ 0.034 & 1.01 $\pm$ 0.16 & -1.142 $\pm$ 0.185 & 3.474 & 206.5 $\pm$ 35.1 & 0   & 192.6 & 1,1,1,1,1\\
J11151597$+$1937266             & $<$0.020          & 0.13 $\pm$ 0.02 & -3.64  $\pm$ 0.12  & 3.230 & 242.2 $\pm$ 37.3 & 0   &  45.2 & 3,4,3,4,5\\
J11432669$-$7804454             & 0.080 $\pm$ 0.027 & 0.85 $\pm$ 0.18 & -1.332 $\pm$ 0.226 & 3.464 & 189.4 $\pm$ 37.7 & 0.4 & 110.0 & 1,1,1,1,1\\
J12071089$-$3230537             & 0.079 $\pm$ 0.009 & 0.38 $\pm$ 0.06 & -1.991 $\pm$ 0.142 & 3.474 & 281.5 $\pm$ 27.4 & 0.7 &  81.7 & 1,1,1,1,1\\
J12073346$-$3932539             & 0.029 $\pm$ 0.003 & 0.28 $\pm$ 0.04 & -2.447 $\pm$ 0.117 & 3.427 & 198.7 $\pm$ 17.5 & 0   &  64.7 & 1,1,1,1,1\\
J15414081$-$3345188             & 0.049 $\pm$ 0.005 & 0.25 $\pm$ 0.01 & -2.445 $\pm$ 0.066 & 3.449 & 273.3 $\pm$ 15.0 & 0.0 & 149.5 & 1,1,1,1,1\\
J15445789$-$3423392             & 0.079 $\pm$ 0.022 & 0.31 $\pm$ 0.04 & -2.157 $\pm$ 0.165 & 3.474 & 311.7 $\pm$ 47.8 & 0   & 154.0 & 1,1,1,1,1\\
J15451851$-$3421246             & 0.047 $\pm$ 0.010 & 0.53 $\pm$ 0.08 & -1.795 $\pm$ 0.164 & 3.449 & 183.9 $\pm$ 24.0 & 0.0 & 151.3 & 1,1,1,1,1\\
J15514032$-$2146103             & 0.111 $\pm$ 0.018 & 0.44 $\pm$ 0.06 & -1.813 $\pm$ 0.139 & 3.489 & 310.1 $\pm$ 32.9 & 0.3 & 140.8 & 1,1,1,1,1\\
J15530132$-$2114135             & 0.111 $\pm$ 0.018 & 0.44 $\pm$ 0.06 & -1.813 $\pm$ 0.139 & 3.489 & 310.1 $\pm$ 32.9 & 0.8 & 143.0 & 1,1,1,1,1\\
J15580252$-$3736026             & 0.099 $\pm$ 0.037 & 0.79 $\pm$ 0.08 & -1.361 $\pm$ 0.150 & 3.474 & 218.6 $\pm$ 42.3 & 0   & 160.0 & 1,1,1,1,1\\
J15582981$-$2310077             & 0.111 $\pm$ 0.018 & 0.44 $\pm$ 0.06 & -1.813 $\pm$ 0.139 & 3.489 & 310.1 $\pm$ 32.9 & 1.0 & 141.1 & 1,1,1,1,1\\
J15591135$-$2338002             & 0.040 $\pm$ 0.002 & 0.29 $\pm$ 0.05 & -2.346 $\pm$ 0.153 & 3.442 & 229.3 $\pm$ 20.6 & 0   & 136.4 & 1,1,1,1,1\\
J15592523$-$4235066             & 0.079 $\pm$ 0.022 & 0.31 $\pm$ 0.04 & -2.157 $\pm$ 0.165 & 3.474 & 311.7 $\pm$ 47.8 & 0.0 & 146.7 & 1,1,1,1,1\\
J16002612$-$4153553             & 0.076 $\pm$ 0.026 & 0.81 $\pm$ 0.16 & -1.375 $\pm$ 0.212 & 3.464 & 189.1 $\pm$ 37.4 & 0.9 & 163.6 & 1,1,1,1,1\\
J16024152$-$2138245             & 0.062 $\pm$ 0.005 & 0.34 $\pm$ 0.06 & -2.129 $\pm$ 0.149 & 3.464 & 263.6 $\pm$ 25.6 & 0.6 & 140.5 & 1,1,1,1,1\\
J16053215$-$1933159$^a$         & 0.10              & 0.57            & -1.569             & 3.490 & 258.6            & 0.8 & 152.3 & 6,6,7,7,7\\
J16060391$-$2056443             & 0.035 $\pm$ 0.005 & 0.29 $\pm$ 0.04 & -2.393 $\pm$ 0.136 & 3.435 & 214.5 $\pm$ 21.3 & 0   & 140.5 & 1,1,1,1,1\\
J16063539$-$2516510             & 0.111 $\pm$ 0.018 & 0.44 $\pm$ 0.06 & -1.813 $\pm$ 0.139 & 3.489 & 310.1 $\pm$ 32.9 & 0   & 138.7 & 1,1,1,1,1\\
J16073773$-$3921388             & 0.073 $\pm$ 0.021 & 0.73 $\pm$ 0.09 & -1.460 $\pm$ 0.145 & 3.464 & 195.2 $\pm$ 30.6 & 0   & 162.4 & 1,1,1,1,1\\
J16080017$-$3902595             & 0.073 $\pm$ 0.021 & 0.74 $\pm$ 0.10 & -1.453 $\pm$ 0.153 & 3.464 & 193.9 $\pm$ 30.8 & 0   & 161.1 & 1,1,1,1,1\\
J16081497$-$3857145             & 0.064 $\pm$ 0.012 & 0.28 $\pm$ 0.02 & -2.286 $\pm$ 0.117 & 3.464 & 295.2 $\pm$ 29.6 & 1.5 & 160.0 & 1,1,1,1,1\\
J16082576$-$3906011             & 0.064 $\pm$ 0.012 & 0.28 $\pm$ 0.02 & -2.286 $\pm$ 0.117 & 3.464 & 295.2 $\pm$ 29.6 & 0   & 160.0 & 1,1,1,1,1\\
J16082751$-$1949047             & 0.062 $\pm$ 0.005 & 0.34 $\pm$ 0.06 & -2.129 $\pm$ 0.149 & 3.464 & 263.6 $\pm$ 25.6 & 0.6 & 145.0 & 1,1,1,1,1\\
J16082847$-$2315103             & 0.026 $\pm$ 0.007 & 0.27 $\pm$ 0.04 & -2.523 $\pm$ 0.183 & 3.418 & 191.6 $\pm$ 29.4 & 0.5 & 169.6 & 1,1,1,1,1\\
J16083081$-$3905488             & 0.104 $\pm$ 0.036 & 0.90 $\pm$ 0.12 & -1.240 $\pm$ 0.171 & 3.474 & 209.9 $\pm$ 38.9 & 0   & 159.8 & 1,1,1,1,1\\
J16083455$-$2211559$^a$         & 0.09              & 0.59            & -1.620             & 3.490 & 241.1            & 0.4 & 139.1 & 6,6,7,7,7\\
J16083733$-$3923109             & 0.032 $\pm$ 0.005 & 0.40 $\pm$ 0.05 & -2.109 $\pm$ 0.150 & 3.435 & 174.6 $\pm$ 17.5 & 0   & 170.7 & 1,1,1,1,1\\
J16085529$-$3848481             & 0.047 $\pm$ 0.010 & 0.54 $\pm$ 0.08 & -1.792 $\pm$ 0.166 & 3.449 & 182.1 $\pm$ 23.6 & 0.0 & 156.9 & 1,1,1,1,1\\
J16085553$-$3902339             & 0.099 $\pm$ 0.037 & 0.80 $\pm$ 0.08 & -1.347 $\pm$ 0.152 & 3.474 & 217.2 $\pm$ 42.0 & 0   & 159.3 & 1,1,1,1,1\\
J16085953$-$3856275             & 0.024 $\pm$ 0.003 & 0.32 $\pm$ 0.03 & -2.368 $\pm$ 0.124 & 3.418 & 169.1 $\pm$ 13.2 & 0   & 155.8 & 1,1,1,1,1\\
J16090002$-$1908368             & 0.111 $\pm$ 0.018 & 0.44 $\pm$ 0.06 & -1.813 $\pm$ 0.139 & 3.489 & 310.1 $\pm$ 32.9 & 0.3 & 136.9 & 1,1,1,1,1\\
J16092697$-$3836269             & 0.106 $\pm$ 0.039 & 0.34 $\pm$ 0.08 & -2.040 $\pm$ 0.272 & 3.489 & 344.7 $\pm$ 75.3 & 2.2 & 158.7 & 1,1,1,1,1\\
J16095361$-$1754474             & 0.111 $\pm$ 0.018 & 0.44 $\pm$ 0.06 & -1.813 $\pm$ 0.139 & 3.489 & 310.1 $\pm$ 32.9 & 0.5 & 158.6 & 1,1,1,1,1\\
J16101857$-$3836125             & 0.099 $\pm$ 0.034 & 0.80 $\pm$ 0.17 & -1.347 $\pm$ 0.223 & 3.474 & 217.2 $\pm$ 43.9 & 0.5 & 158.1 & 1,1,1,1,1\\
J16102819$-$1910444$^a$         & 0.08              & 0.48            & -1.569             & 3.490 & 252.1            & 1.0 & 151.7 & 6,6,7,7,7\\
J16104636$-$1840598             & 0.111 $\pm$ 0.018 & 0.44 $\pm$ 0.06 & -1.813 $\pm$ 0.139 & 3.489 & 310.1 $\pm$ 32.9 & 1.2 & 141.4 & 1,1,1,1,1\\
J16115979$-$3823383             & 0.100 $\pm$ 0.037 & 0.82 $\pm$ 0.09 & -1.331 $\pm$ 0.159 & 3.474 & 215.6 $\pm$ 41.6 & 0.0 & 163.9 & 1,1,1,1,1\\
J16134410$-$3736462             & 0.079 $\pm$ 0.022 & 0.31 $\pm$ 0.04 & -2.157 $\pm$ 0.165 & 3.474 & 311.7 $\pm$ 47.8 & 0.6 & 158.5 & 1,1,1,1,1\\
J16135434$-$2320342             & 0.111 $\pm$ 0.018 & 0.44 $\pm$ 0.06 & -1.813 $\pm$ 0.139 & 3.489 & 310.1 $\pm$ 32.9 & 0.3 & 145.0 & 1,1,1,1,1\\
J16181904$-$2028479             & 0.079 $\pm$ 0.009 & 0.38 $\pm$ 0.06 & -1.991 $\pm$ 0.142 & 3.474 & 281.5 $\pm$ 27.4 & 1.6 & 139.7 & 1,1,1,1,1\\
J16262189$-$2444397             & 0.046 $\pm$ 0.003 & 0.31 $\pm$ 0.05 & -2.276 $\pm$ 0.157 & 3.449 & 237.8 $\pm$ 20.7 & 0.6 & 140.0 & 1,1,1,1,1\\
J16272658$-$2425543$^a$         & 0.033             & 0.4$^b$         & -1.921             & 3.435 & 177.3            & 1.0 & 137.3 & 7,--,7,7,7\\
\enddata
\tablecomments{
Column~1: the 2MASS designations.
Column~2: object masses from the literature (1st reference entries in column~9).
Column~3: object radii from the literature (2nd reference entries in column~9).
Column~4: object luminosities from the literature (3nd reference entries in column~9).
Column~5: effective temperatures from the literature (4th reference entries in column~9).
Column~6: free-fall velocities from infinity: $v_{\rm ff}=\sqrt{2GM_*/R_*}$, where $M_*$ is an object mass in column~2, $R_*$ is an object radius in column~3, and $G$ is the gravitational constant.
Column~7: visual extinctions from the literature (5th reference entries in column~9).
Column~8: GAIA distances \citep{Gaia2016,Gaia2023}.
Column~9: the references for the $M_*$, $R_*$, $L_*$, $T_{\rm eff}$, and visual extinctions, respectively:
(1) \citet{Betti2023Survey};
(2) \citet{Herczeg2008ff};
(3) this study (refer to \$~\ref{secA:2m1115};
(4) \citet{Theissen2018-2m1115};
(5) \citet{Theissen2017-2m1115};
(6) \citet{Fang2023USco};
(7) \citet{Almendros-Abad2024YBD}.
}
\tablenotetext{a}{Errors in columns 2 to 4 were not provided in the literature.}\vspace{-0.2cm} 
\tablenotetext{b}{An estimate from the COND model at 1~Myr \citep{Baraffe2003COND}.}\vspace{-0.2cm} 
\end{deluxetable*}
\end{startlongtable}

\begin{longrotatetable}
\begin{deluxetable*}{lccccccccc}
\tablecaption{Accretion properties of targets.\label{tabA:flux}}
\tablewidth{700pt}
\tabletypesize{\scriptsize}
\tablehead{
\colhead{2MASS Name} & \colhead{H$\beta$}   & \colhead{H$\beta$/H$\gamma$} & \colhead{H$\beta$/H8} & 
\colhead{log ($L_{\rm acc,A17}/L_\sun$)} & \colhead{log ($L_{\rm acc,A21}/L_\sun$)} &  
\colhead{log ($\dot{M}_{\rm A17}/(M_\sun/{\rm yr}$)} & \colhead{log ($\dot{M}_{\rm A21}/(M_\sun/{\rm yr}$)} & \colhead{Em?} & \colhead{Refs} \\ 
\colhead{}           & \colhead{($10^{-7} L_\sun$)} & \colhead{} & \colhead{} & 
\colhead{(dex)} & \colhead{(dex)} & 
\colhead{(dex)} & \colhead{(dex)} & \colhead{} & \colhead{}
} 
\decimalcolnumbers
\startdata
J04141188$+$2811535              &  387.68             &  1.63            &  3.48            & -2.44 $\pm$ 0.30 & -2.37 $\pm$ 0.12 &  -8.8 $\pm$ 0.3 &  -8.8 $\pm$ 0.2 & Flow & 1\\
J04183030$+$2743208              &   12.73             &  1.98            &  5.95            & -4.13 $\pm$ 0.30 & -3.66 $\pm$ 0.12 & -10.6 $\pm$ 0.3 & -10.1 $\pm$ 0.2 & Chr & 1\\
J04262939$+$2624137              &   25.10 $\pm$  1.00 &  1.72 $\pm$ 0.16 &  2.96 $\pm$ 0.49 & -3.79 $\pm$ 0.30 & -3.40 $\pm$ 0.12 & -10.3 $\pm$ 0.3 &  -9.9 $\pm$ 0.1 & Flow & 2\\
J04321606$+$1812464              &   79.04             &  2.31            &  8.98            & -3.23 $\pm$ 0.30 & -2.97 $\pm$ 0.12 &  -9.7 $\pm$ 0.3 &  -9.4 $\pm$ 0.1 & Flow& 1\\
J04321786$+$2422149              &    4.21             &  2.00            & 10.20            & -4.68 $\pm$ 0.30 & -4.08 $\pm$ 0.12 & -11.2 $\pm$ 0.3 & -10.6 $\pm$ 0.1 & Chr & 1\\
J04322210$+$1827426 (2006-11-23) &  309.35             &  1.35            &  2.88            & -2.55 $\pm$ 0.30 & -2.45 $\pm$ 0.12 &  -9.1 $\pm$ 0.3 &  -9.0 $\pm$ 0.1 & Shock & 1\\
J04322210$+$1827426 (2015-10-25) &   39.45 $\pm$  0.70 &  1.54 $\pm$ 0.06 &  3.48 $\pm$ 0.36 & -3.57 $\pm$ 0.30 & -3.23 $\pm$ 0.12 & -10.1 $\pm$ 0.3 &  -9.8 $\pm$ 0.1 & Flow & 2\\
J04362151$+$2351165              &   57.39             &  1.07            &  2.22            & -3.38 $\pm$ 0.30 & -3.09 $\pm$ 0.12 &  -9.9 $\pm$ 0.3 &  -9.6 $\pm$ 0.2 & Shock & 1\\
J04382134$+$2609137              &  572.38             &  1.94            &  2.87            & -2.25 $\pm$ 0.30 & -2.22 $\pm$ 0.12 &  -8.7 $\pm$ 0.3 &  -8.7 $\pm$ 0.2 & N/A & 1\\
J04390163$+$2336029 (2006-11-23) &   78.11             &  1.25            &  3.29            & -3.23 $\pm$ 0.30 & -2.97 $\pm$ 0.12 &  -9.8 $\pm$ 0.3 &  -9.5 $\pm$ 0.2 & Shock & 1\\
J04390163$+$2336029 (2015-11-22) &   36.95 $\pm$  0.27 &  1.42 $\pm$ 0.02 &  2.57 $\pm$ 0.06 & -3.60 $\pm$ 0.30 & -3.26 $\pm$ 0.12 & -10.1 $\pm$ 0.3 &  -9.8 $\pm$ 0.2 & Flow & 2\\
J04390396$+$2544264              &   17.55             &  1.93            &  5.83            & -3.97 $\pm$ 0.30 & -3.54 $\pm$ 0.12 & -10.4 $\pm$ 0.3 &  -9.9 $\pm$ 0.2 & Flow & 1\\
J04414825$+$2534304              &   27.51             &  1.61            &  4.09            & -3.75 $\pm$ 0.30 & -3.37 $\pm$ 0.12 & -10.1 $\pm$ 0.3 &  -9.8 $\pm$ 0.1 & Flow & 1\\
J04432023$+$2940060              &   50.06             &  2.43            &  5.52            & -3.45 $\pm$ 0.30 & -3.14 $\pm$ 0.12 & -10.0 $\pm$ 0.3 &  -9.7 $\pm$ 0.1 & Flow & 1\\
J04442713$+$2512164              &   21.46             &  2.27            & 11.00            & -3.87 $\pm$ 0.30 & -3.46 $\pm$ 0.12 & -10.3 $\pm$ 0.3 &  -9.9 $\pm$ 0.2 & both & 1\\
J04554757$+$3028077              &  112.01             &  1.47            &  4.04            & -3.05 $\pm$ 0.30 & -2.84 $\pm$ 0.12 &  -9.7 $\pm$ 0.3 &  -9.5 $\pm$ 0.2 & Shock & 1\\
J05180285$+$2327127              &  148.38             &  1.00            &  1.79            & -2.91 $\pm$ 0.30 & -2.73 $\pm$ 0.12 &  -9.7 $\pm$ 0.3 &  -9.5 $\pm$ 0.2 & Shock & 1\\
J05382358$-$0220475              &  206.12 $\pm$  0.90 &  1.33 $\pm$ 0.01 &  2.00 $\pm$ 0.02 & -2.75 $\pm$ 0.30 & -2.61 $\pm$ 0.12 &  -9.2 $\pm$ 0.3 &  -9.1 $\pm$ 0.2 & Flow & 2\\
J05382543$-$0242412              &   87.81 $\pm$  0.68 &  2.07 $\pm$ 0.03 &  3.88 $\pm$ 0.12 & -3.17 $\pm$ 0.30 & -2.93 $\pm$ 0.12 &  -9.6 $\pm$ 0.3 &  -9.3 $\pm$ 0.1 & Flow & 2\\
J05384818$-$0244007              &   55.61 $\pm$  1.65 &  1.40 $\pm$ 0.07 &  2.65 $\pm$ 0.28 & -3.40 $\pm$ 0.30 & -3.10 $\pm$ 0.12 & -10.0 $\pm$ 0.3 &  -9.7 $\pm$ 0.2 & Both & 2\\
J05395173$-$0222472              &  732.42 $\pm$  3.48 &  2.19 $\pm$ 0.03 &  3.67 $\pm$ 0.09 & -2.12 $\pm$ 0.30 & -2.13 $\pm$ 0.12 &  -8.7 $\pm$ 0.3 &  -8.7 $\pm$ 0.2 & Flow & 2\\
J08361073$-$7908184              &    3.70 $\pm$  0.04 &  2.20 $\pm$ 0.05 &  5.03 $\pm$ 0.31 & -4.74 $\pm$ 0.30 & -4.13 $\pm$ 0.12 & -11.5 $\pm$ 0.3 & -10.9 $\pm$ 0.1 & Flow & 2\\
J08385150$-$7916137              &    8.52 $\pm$  0.17 &  2.18 $\pm$ 0.10 &  4.61 $\pm$ 0.66 & -4.33 $\pm$ 0.30 & -3.81 $\pm$ 0.12 & -11.1 $\pm$ 0.3 & -10.6 $\pm$ 0.1 & Flow & 2\\
J08413030$-$7853064              &    5.26 $\pm$  0.08 &  1.90 $\pm$ 0.06 &  5.02 $\pm$ 0.49 & -4.57 $\pm$ 0.30 & -3.99 $\pm$ 0.12 & -11.4 $\pm$ 0.3 & -10.8 $\pm$ 0.1 & Flow & 2\\
J08422710$-$7857479 (2010-01-19) &   50.28 $\pm$  0.37 &  1.45 $\pm$ 0.02 &  2.60 $\pm$ 0.07 & -3.45 $\pm$ 0.30 & -3.14 $\pm$ 0.12 & -10.4 $\pm$ 0.3 & -10.1 $\pm$ 0.1 & Flow & 2\\
J08422710$-$7857479 (2018-06-18) &   14.56 $\pm$  0.06 &  1.65 $\pm$ 0.01 &  3.65 $\pm$ 0.07 & -4.06 $\pm$ 0.30 & -3.61 $\pm$ 0.12 & -11.0 $\pm$ 0.3 & -10.5 $\pm$ 0.1 & Flow & 2\\
J08422710$-$7857479 (2022-01-28) &   20.92 $\pm$  0.11 &  1.96 $\pm$ 0.02 &  4.79 $\pm$ 0.11 & -3.88 $\pm$ 0.30 & -3.47 $\pm$ 0.12 & -10.8 $\pm$ 0.3 & -10.4 $\pm$ 0.1 & Flow & 2\\
J08440915$-$7833457 (2010-01-18) &   26.61 $\pm$  0.13 &  1.39 $\pm$ 0.01 &  2.13 $\pm$ 0.04 & -3.77 $\pm$ 0.30 & -3.38 $\pm$ 0.12 & -10.5 $\pm$ 0.3 & -10.1 $\pm$ 0.1 & Flow & 2\\
J08440915$-$7833457 (2021-04-27) &   12.72 $\pm$  0.01 &  1.32 $\pm$ 0.00 &  2.48 $\pm$ 0.01 & -4.13 $\pm$ 0.30 & -3.66 $\pm$ 0.12 & -10.9 $\pm$ 0.3 & -10.4 $\pm$ 0.1 & Shock & 2\\
J08440915$-$7833457 (2021-05-01) &   18.04 $\pm$  0.06 &  1.31 $\pm$ 0.01 &  2.43 $\pm$ 0.03 & -3.96 $\pm$ 0.30 & -3.53 $\pm$ 0.12 & -10.7 $\pm$ 0.3 & -10.3 $\pm$ 0.1 & Shock & 2\\
J08441637$-$7859080 (2010-01-19) &   37.99 $\pm$  0.33 &  1.53 $\pm$ 0.02 &  3.02 $\pm$ 0.11 & -3.59 $\pm$ 0.30 & -3.25 $\pm$ 0.12 & -10.5 $\pm$ 0.3 & -10.2 $\pm$ 0.1 & Flow & 2\\
J08441637$-$7859080 (2022-01-26) &   26.84 $\pm$  0.08 &  1.56 $\pm$ 0.01 &  2.69 $\pm$ 0.02 & -3.76 $\pm$ 0.30 & -3.38 $\pm$ 0.12 & -10.7 $\pm$ 0.3 & -10.3 $\pm$ 0.1 & Flow & 2\\
J10561638$-$7630530              &    6.75 $\pm$  0.24 &  2.00 $\pm$ 0.19 &  4.48 $\pm$ 1.20 & -4.44 $\pm$ 0.30 & -3.90 $\pm$ 0.12 & -10.9 $\pm$ 0.3 & -10.4 $\pm$ 0.1 & Chr & 2\\
J11020983$-$3430355              &    1.29 $\pm$  0.09 &  1.42 $\pm$ 0.16 &  2.74 $\pm$ 0.84 & -5.26 $\pm$ 0.30 & -4.52 $\pm$ 0.12 & -11.7 $\pm$ 0.3 & -11.0 $\pm$ 0.1 & Both & 3\\
J11064180$-$7635489 (2010-01-18) &   92.83 $\pm$  0.69 &  1.54 $\pm$ 0.02 &  2.71 $\pm$ 0.08 & -3.15 $\pm$ 0.30 & -2.91 $\pm$ 0.12 &  -9.7 $\pm$ 0.3 &  -9.5 $\pm$ 0.1 & Flow & 2\\
J11064180$-$7635489 (2021-06-08) &   84.62 $\pm$  0.26 &  1.95 $\pm$ 0.01 &  3.89 $\pm$ 0.05 & -3.19 $\pm$ 0.30 & -2.94 $\pm$ 0.12 &  -9.7 $\pm$ 0.3 &  -9.5 $\pm$ 0.1 & N/A & 2\\
J11065939$-$7530559              &   17.71 $\pm$  0.16 &  1.26 $\pm$ 0.02 &  2.18 $\pm$ 0.08 & -3.97 $\pm$ 0.30 & -3.53 $\pm$ 0.12 & -10.4 $\pm$ 0.3 & -10.0 $\pm$ 0.2 & Both & 2\\
J11081850$-$7730408              &    3.88 $\pm$  0.11 &  1.37 $\pm$ 0.07 &  3.38 $\pm$ 0.52 & -4.72 $\pm$ 0.30 & -4.11 $\pm$ 0.12 & -11.1 $\pm$ 0.3 & -10.5 $\pm$ 0.2 & Chr & 2\\
J11082238$-$7730277              &   18.98 $\pm$  0.27 &  1.16 $\pm$ 0.03 &  2.34 $\pm$ 0.16 & -3.93 $\pm$ 0.30 & -3.51 $\pm$ 0.12 & -10.4 $\pm$ 0.3 & -10.0 $\pm$ 0.1 & Shock & 2\\
J11083952$-$7734166              &   13.49 $\pm$  0.09 &  3.15 $\pm$ 0.07 &  9.54 $\pm$ 0.89 & -4.10 $\pm$ 0.30 & -3.64 $\pm$ 0.12 & -10.6 $\pm$ 0.3 & -10.1 $\pm$ 0.2 & Flow & 2\\
J11085090$-$7625135              &   47.72 $\pm$  0.63 &  1.28 $\pm$ 0.03 &  2.64 $\pm$ 0.20 & -3.48 $\pm$ 0.30 & -3.16 $\pm$ 0.12 &  -9.9 $\pm$ 0.3 &  -9.6 $\pm$ 0.2 & Shock & 2\\
J11120984$-$7634366 (2015-04-14) &   84.09 $\pm$  0.38 &  1.56 $\pm$ 0.01 &  2.84 $\pm$ 0.05 & -3.20 $\pm$ 0.30 & -2.95 $\pm$ 0.12 &  -9.7 $\pm$ 0.3 &  -9.5 $\pm$ 0.2 & Flow & 2\\
J11120984$-$7634366 (2021-06-08) &   58.34 $\pm$  0.22 &  1.77 $\pm$ 0.01 &  3.72 $\pm$ 0.07 & -3.38 $\pm$ 0.30 & -3.08 $\pm$ 0.12 &  -9.9 $\pm$ 0.3 &  -9.6 $\pm$ 0.2 & Flow & 2\\
J11151597$+$1937266              &    1.45 $\pm$  0.02 &  1.36 $\pm$ 0.03 &  2.79 $\pm$ 0.11 & -5.21 $\pm$ 0.30 & -4.48 $\pm$ 0.12 & -11.9 $\pm$ 0.3 & -11.2 $\pm$ 0.1 & Shock & 4\\
J11432669$-$7804454 (2015-01-04) &   58.72 $\pm$  0.42 &  1.44 $\pm$ 0.02 &  2.36 $\pm$ 0.08 & -3.37 $\pm$ 0.30 & -3.08 $\pm$ 0.12 &  -9.8 $\pm$ 0.3 &  -9.6 $\pm$ 0.2 & Flow & 2\\
J11432669$-$7804454 (2015-04-20) &  146.37 $\pm$  0.22 &  1.44 $\pm$ 0.00 &  1.88 $\pm$ 0.01 & -2.92 $\pm$ 0.30 & -2.74 $\pm$ 0.12 &  -9.4 $\pm$ 0.3 &  -9.2 $\pm$ 0.2 & Flow & 2\\
J12071089$-$3230537              &  272.50 $\pm$  6.30 &  1.57 $\pm$ 0.09 &  3.07 $\pm$ 0.19 & -2.61 $\pm$ 0.30 & -2.50 $\pm$ 0.12 &  -9.4 $\pm$ 0.3 &  -9.3 $\pm$ 0.1 & Flow & 3\\
J12073346$-$3932539 (2006-11-23) &    2.49             &  1.60            &  3.65            & -4.94 $\pm$ 0.30 & -4.28 $\pm$ 0.12 & -11.4 $\pm$ 0.3 & -10.8 $\pm$ 0.1 & Flow & 1\\
J12073346$-$3932539 (2007-02-07) &    3.54             &  1.96            &  5.35            & -4.76 $\pm$ 0.30 & -4.14 $\pm$ 0.12 & -11.3 $\pm$ 0.3 & -10.7 $\pm$ 0.1 & Flow & 1\\
J12073346$-$3932539 (2010-03-23)       &    2.80 $\pm$  0.20 &  0.93 $\pm$ 0.11 &  1.75 $\pm$ 0.35 & -4.88 $\pm$ 0.30 & -4.23 $\pm$ 0.12 & -11.4 $\pm$ 0.3 & -10.7 $\pm$ 0.1 & Shock & 3\\
J12073346$-$3932539 (2012-04-19)       &    2.46 $\pm$  0.09 &  1.58 $\pm$ 0.07 &  2.96 $\pm$ 0.24 & -4.94 $\pm$ 0.30 & -4.28 $\pm$ 0.12 & -11.5 $\pm$ 0.3 & -10.8 $\pm$ 0.1 & Flow & 3\\
J15414081$-$3345188              &    4.66 $\pm$  0.42 &  1.72 $\pm$ 0.21 &  2.53 $\pm$ 0.39 & -4.63 $\pm$ 0.30 & -4.04 $\pm$ 0.12 & -11.4 $\pm$ 0.3 & -10.8 $\pm$ 0.1 & Flow & 5\\
J15445789$-$3423392              &   14.10 $\pm$  0.40 &  1.59 $\pm$ 0.07 &  3.39 $\pm$ 0.30 & -4.08 $\pm$ 0.30 & -3.62 $\pm$ 0.12 & -11.0 $\pm$ 0.3 & -10.5 $\pm$ 0.2 & Flow & 6\\
J15451851$-$3421246              &   34.30 $\pm$  1.10 &  2.52 $\pm$ 0.14 &  6.45 $\pm$ 0.76 & -3.64 $\pm$ 0.30 & -3.28 $\pm$ 0.12 & -10.1 $\pm$ 0.3 &  -9.7 $\pm$ 0.2 & Flow & 5\\
J15514032$-$2146103              &   11.14 $\pm$  0.11 &  1.99 $\pm$ 0.04 &  4.55 $\pm$ 0.23 & -4.20 $\pm$ 0.30 & -3.71 $\pm$ 0.12 & -11.1 $\pm$ 0.3 & -10.6 $\pm$ 0.2 & Flow & 2\\
J15530132$-$2114135              &   57.28 $\pm$  0.20 &  1.32 $\pm$ 0.01 &  2.22 $\pm$ 0.02 & -3.39 $\pm$ 0.30 & -3.09 $\pm$ 0.12 & -10.3 $\pm$ 0.3 & -10.0 $\pm$ 0.2 & Both & 2\\
J15580252$-$3736026 (2012-04-18) &  194.11 $\pm$  0.19 &  1.88 $\pm$ 0.01 &  2.93 $\pm$ 0.03 & -2.78 $\pm$ 0.30 & -2.63 $\pm$ 0.12 &  -9.4 $\pm$ 0.3 &  -9.2 $\pm$ 0.2 & N/A & 6\\
J15580252$-$3736026 (2022-05-11) &   45.66 $\pm$  0.25 &  1.26 $\pm$ 0.01 &  2.17 $\pm$ 0.04 & -3.50 $\pm$ 0.30 & -3.18 $\pm$ 0.12 & -10.1 $\pm$ 0.3 &  -9.8 $\pm$ 0.2 & Both & 2\\
J15582981$-$2310077              &  607.26 $\pm$  0.54 &  1.30 $\pm$ 0.00 &  2.13 $\pm$ 0.01 & -2.22 $\pm$ 0.30 & -2.20 $\pm$ 0.12 &  -9.1 $\pm$ 0.3 &  -9.1 $\pm$ 0.2 & Both & 2\\
J15591135$-$2338002              &    1.93 $\pm$  0.04 &  1.34 $\pm$ 0.04 &  3.28 $\pm$ 0.24 & -5.06 $\pm$ 0.30 & -4.37 $\pm$ 0.12 & -11.7 $\pm$ 0.3 & -11.0 $\pm$ 0.1 & Shock & 2\\
J15592523$-$4235066              &    6.69 $\pm$  0.57 &  1.60 $\pm$ 0.14 &  4.16 $\pm$ 0.90 & -4.45 $\pm$ 0.30 & -3.90 $\pm$ 0.12 & -11.4 $\pm$ 0.3 & -10.8 $\pm$ 0.2 & Both & 5\\
J16002612$-$4153553              &   67.50 $\pm$  9.10 &  1.35 $\pm$ 0.26 &  2.26 $\pm$ 0.55 & -3.30 $\pm$ 0.30 & -3.03 $\pm$ 0.12 &  -9.8 $\pm$ 0.3 &  -9.5 $\pm$ 0.2 & Both & 5\\
J16024152$-$2138245              &  101.42 $\pm$  1.01 &  1.56 $\pm$ 0.03 &  3.21 $\pm$ 0.13 & -3.10 $\pm$ 0.30 & -2.87 $\pm$ 0.12 &  -9.9 $\pm$ 0.3 &  -9.6 $\pm$ 0.1 & Flow & 2\\
J16053215$-$1933159              &   48.20 $\pm$  0.50 &  1.23 $\pm$ 0.02 &  2.30 $\pm$ 0.07 & -3.47 $\pm$ 0.30 & -3.16 $\pm$ 0.12 & -10.2 $\pm$ 0.3 &  -9.9 $\pm$ 0.1 & Shock & 2\\
J16060391$-$2056443              &    4.37 $\pm$  0.04 &  2.33 $\pm$ 0.05 &  7.33 $\pm$ 0.54 & -4.66 $\pm$ 0.30 & -4.06 $\pm$ 0.12 & -11.2 $\pm$ 0.3 & -10.6 $\pm$ 0.1 & Flow & 2\\
J16063539$-$2516510              &    7.94 $\pm$  0.07 &  1.83 $\pm$ 0.03 &  3.80 $\pm$ 0.17 & -4.36 $\pm$ 0.30 & -3.84 $\pm$ 0.12 & -11.3 $\pm$ 0.3 & -10.7 $\pm$ 0.2 & Chr & 2\\
J16073773$-$3921388 (2010-04-06) &   67.36 $\pm$  0.18 &  2.02 $\pm$ 0.01 &  3.55 $\pm$ 0.04 & -3.31 $\pm$ 0.30 & -3.03 $\pm$ 0.12 &  -9.8 $\pm$ 0.3 &  -9.5 $\pm$ 0.2 & N/A & 2\\
J16073773$-$3921388 (2012-04-18) &   55.49 $\pm$  0.16 &  1.74 $\pm$ 0.01 &  3.02 $\pm$ 0.03 & -3.40 $\pm$ 0.30 & -3.10 $\pm$ 0.12 &  -9.9 $\pm$ 0.3 &  -9.6 $\pm$ 0.2 & Flow & 2\\
J16073773$-$3921388 (2012-04-19) &   29.90 $\pm$  0.13 &  1.99 $\pm$ 0.02 &  3.48 $\pm$ 0.07 & -3.71 $\pm$ 0.30 & -3.34 $\pm$ 0.12 & -10.2 $\pm$ 0.3 &  -9.8 $\pm$ 0.2 & N/A & 2\\
J16080017$-$3902595              &   12.70 $\pm$  0.60 &  1.83 $\pm$ 0.19 &  3.62 $\pm$ 0.57 & -4.13 $\pm$ 0.30 & -3.66 $\pm$ 0.12 & -10.6 $\pm$ 0.3 & -10.2 $\pm$ 0.2 & Flow & 6\\
J16081497$-$3857145              &   29.90 $\pm$  2.60 &  1.59 $\pm$ 0.23 &  3.16 $\pm$ 0.42 & -3.71 $\pm$ 0.30 & -3.34 $\pm$ 0.12 & -10.6 $\pm$ 0.3 & -10.2 $\pm$ 0.1 & Both & 5\\
J16082576$-$3906011 (2011-04-23) &  118.52 $\pm$  0.47 &  1.50 $\pm$ 0.01 &  3.01 $\pm$ 0.05 & -3.03 $\pm$ 0.30 & -2.82 $\pm$ 0.12 &  -9.9 $\pm$ 0.3 &  -9.7 $\pm$ 0.1 & Flow & 6\\
J16082576$-$3906011 (2022-06-24) &  152.06 $\pm$  0.30 &  1.54 $\pm$ 0.01 &  2.86 $\pm$ 0.02 & -2.90 $\pm$ 0.30 & -2.72 $\pm$ 0.12 &  -9.8 $\pm$ 0.3 &  -9.6 $\pm$ 0.1 & Flow & 2\\
J16082751$-$1949047              &   28.06 $\pm$  0.21 &  1.41 $\pm$ 0.02 &  2.64 $\pm$ 0.08 & -3.74 $\pm$ 0.30 & -3.36 $\pm$ 0.12 & -10.5 $\pm$ 0.3 & -10.1 $\pm$ 0.1 & Flow & 2\\
J16082847$-$2315103              &    3.12 $\pm$  0.17 &  1.62 $\pm$ 0.19 &  1.51 $\pm$ 0.25 & -4.83 $\pm$ 0.30 & -4.19 $\pm$ 0.12 & -11.3 $\pm$ 0.3 & -10.7 $\pm$ 0.2 & N/A & 2\\
J16083081$-$3905488 (2010-04-06) &   60.72 $\pm$  0.38 &  1.44 $\pm$ 0.02 &  2.71 $\pm$ 0.07 & -3.36 $\pm$ 0.30 & -3.07 $\pm$ 0.12 &  -9.9 $\pm$ 0.3 &  -9.6 $\pm$ 0.2 & Flow & 6\\
J16083081$-$3905488 (2022-06-24) &   53.83 $\pm$  0.25 &  1.31 $\pm$ 0.01 &  2.36 $\pm$ 0.04 & -3.42 $\pm$ 0.30 & -3.11 $\pm$ 0.12 & -10.0 $\pm$ 0.3 &  -9.7 $\pm$ 0.2 & Shock & 2\\
J16083455$-$2211559              &    4.98 $\pm$  0.21 &  1.27 $\pm$ 0.10 &  2.77 $\pm$ 0.45 & -4.60 $\pm$ 0.30 & -4.01 $\pm$ 0.12 & -11.3 $\pm$ 0.3 & -10.7 $\pm$ 0.1 & Chr & 2\\
J16083733$-$3923109              &    2.77 $\pm$  0.09 &  2.52 $\pm$ 0.24 &  7.49 $\pm$ 2.44 & -4.89 $\pm$ 0.30 & -4.24 $\pm$ 0.12 & -11.3 $\pm$ 0.3 & -10.6 $\pm$ 0.1 & Flow & 6\\
J16085529$-$3848481              &   10.60 $\pm$  0.50 &  1.69 $\pm$ 0.14 &  3.76 $\pm$ 0.39 & -4.22 $\pm$ 0.30 & -3.73 $\pm$ 0.12 & -10.7 $\pm$ 0.3 & -10.2 $\pm$ 0.2 & Flow & 5\\
J16085553$-$3902339 (2012-04-18) &   52.59 $\pm$  0.27 &  2.02 $\pm$ 0.02 &  4.11 $\pm$ 0.09 & -3.43 $\pm$ 0.30 & -3.12 $\pm$ 0.12 & -10.0 $\pm$ 0.3 &  -9.7 $\pm$ 0.2 & Flow & 5\\
J16085553$-$3902339 (2022-07-24) &   81.67 $\pm$  0.29 &  1.63 $\pm$ 0.01 &  3.19 $\pm$ 0.04 & -3.21 $\pm$ 0.30 & -2.96 $\pm$ 0.12 &  -9.8 $\pm$ 0.3 &  -9.5 $\pm$ 0.2 & Flow & 2\\
J16085953$-$3856275              &    6.38 $\pm$  0.16 &  1.52 $\pm$ 0.07 &  2.91 $\pm$ 0.20 & -4.47 $\pm$ 0.30 & -3.92 $\pm$ 0.12 & -10.8 $\pm$ 0.3 & -10.3 $\pm$ 0.1 & Flow & 6\\
J16090002$-$1908368              &   16.54 $\pm$  0.11 &  1.77 $\pm$ 0.02 &  3.92 $\pm$ 0.12 & -4.00 $\pm$ 0.30 & -3.56 $\pm$ 0.12 & -10.9 $\pm$ 0.3 & -10.5 $\pm$ 0.2 & Flow & 2\\
J16092697$-$3836269              & 1634.00 $\pm$ 71.00 &  1.20 $\pm$ 0.09 &  1.64 $\pm$ 0.11 & -1.73 $\pm$ 0.30 & -1.82 $\pm$ 0.12 &  -8.7 $\pm$ 0.3 &  -8.8 $\pm$ 0.2 & Flow & 5\\
J16095361$-$1754474              &   23.62 $\pm$  0.15 &  1.53 $\pm$ 0.02 &  2.99 $\pm$ 0.08 & -3.82 $\pm$ 0.30 & -3.43 $\pm$ 0.12 & -10.7 $\pm$ 0.3 & -10.3 $\pm$ 0.2 & Flow & 2\\
J16101857$-$3836125              &   19.80 $\pm$  1.90 &  1.42 $\pm$ 0.21 &  2.57 $\pm$ 0.50 & -3.91 $\pm$ 0.30 & -3.49 $\pm$ 0.12 & -10.5 $\pm$ 0.3 & -10.1 $\pm$ 0.2 & Both & 5\\
J16102819$-$1910444              &    9.78 $\pm$  0.26 &  1.42 $\pm$ 0.08 &  3.09 $\pm$ 0.39 & -4.26 $\pm$ 0.30 & -3.76 $\pm$ 0.12 & -11.0 $\pm$ 0.3 & -10.5 $\pm$ 0.1 & Chr & 2\\
J16104636$-$1840598              &   14.84 $\pm$  0.23 &  1.58 $\pm$ 0.05 &  3.12 $\pm$ 0.28 & -4.05 $\pm$ 0.30 & -3.60 $\pm$ 0.12 & -11.0 $\pm$ 0.3 & -10.5 $\pm$ 0.2 & Flow & 2\\
J16115979$-$3823383              &   72.60 $\pm$  2.20 &  1.40 $\pm$ 0.05 &  2.99 $\pm$ 0.23 & -3.27 $\pm$ 0.30 & -3.00 $\pm$ 0.12 &  -9.9 $\pm$ 0.3 &  -9.6 $\pm$ 0.2 & Both & 5\\
J16134410$-$3736462 (2015-06-26) &  319.67 $\pm$  0.81 &  1.19 $\pm$ 0.01 &  1.85 $\pm$ 0.01 & -2.53 $\pm$ 0.30 & -2.44 $\pm$ 0.12 &  -9.4 $\pm$ 0.3 &  -9.3 $\pm$ 0.2 & Both & 5\\
J16134410$-$3736462 (2022-05-03) &  283.87 $\pm$  0.46 &  1.09 $\pm$ 0.00 &  1.71 $\pm$ 0.01 & -2.59 $\pm$ 0.30 & -2.49 $\pm$ 0.12 &  -9.5 $\pm$ 0.3 &  -9.4 $\pm$ 0.2 & Shock & 2\\
J16135434$-$2320342              &  537.73 $\pm$  0.81 &  1.81 $\pm$ 0.01 &  3.53 $\pm$ 0.02 & -2.28 $\pm$ 0.30 & -2.24 $\pm$ 0.12 &  -9.2 $\pm$ 0.3 &  -9.1 $\pm$ 0.2 & Flow & 2\\
J16181904$-$2028479              &   25.89 $\pm$  0.31 &  1.71 $\pm$ 0.05 &  3.38 $\pm$ 0.27 & -3.78 $\pm$ 0.30 & -3.39 $\pm$ 0.12 & -10.6 $\pm$ 0.3 & -10.2 $\pm$ 0.1 & Flow & 2\\
J16262189$-$2444397              &    8.07 $\pm$  0.08 &  1.66 $\pm$ 0.03 &  2.97 $\pm$ 0.16 & -4.36 $\pm$ 0.30 & -3.83 $\pm$ 0.12 & -11.0 $\pm$ 0.3 & -10.5 $\pm$ 0.1 & Flow & 2\\
J16272658$-$2425543              &   13.60 $\pm$  0.20 &  1.74 $\pm$ 0.07 &  3.95 $\pm$ 0.37 & -4.10 $\pm$ 0.30 & -3.63 $\pm$ 0.12 & -10.5 $\pm$ 0.3 & -10.0 $\pm$ 0.1 & Flow & 2\\
\enddata
\tablecomments{
Column~1: the 2MASS designations.
Column~2: extinction-corrected H$\beta$ luminosities from the literature (1st reference entries in column~10).
Column~3: line ratios of H$\beta$/H$\gamma$ from the literature (1st reference entries in column~9).
Column~4: line ratios of H$\beta$/H8 from the literature (1st reference entries in column~9).
Column~5: accretion luminosities converted from H$\beta$ luminosities in column~2 using the $L_{\rm acc}$-$L_{\rm line}$ relations in \citet{alca2017}. The errors are assumed to standard deviation from linear in \citet{alca2017}.
Column~6: same with Column~5, but \citet{Aoyama2021conversion}.
Column~7: mass accretion rates estimated by eq.~\ref{eq:mdot} and accretion luminosities in Column~5.
Column~8: same with Column~7, but accretion luminosities in Column~6.
Column~9: emission mechanisms categorized in \S~\ref{subsec:plot}. The `Shock', `Flow', `Both', `N/A', and `Chr' are the shock/flow/both models and none of them, and the chromospheric dominated emission, respectively.
Column~10: the references for the extinction-corrected line luminosities of H$\beta$, H$\gamma$, and H8 for Columns~2 to 4:
(1) \citet{Herczeg2008ff};
(2) this study using VLT/Xshooter archival data;
(3) \citet{Venuti2019TWA};
(4) this study using SDSS archival data;
(5) \citet{alca2017};
(6) \citet{Alcala2014XSOOTER}.
}
\tablenotetext{a}{Errors in columns~2 to 4 were not provided in the literature.}\vspace{-0.2cm} 
\end{deluxetable*}
\end{longrotatetable}

\begin{figure}[htbp]
    \includegraphics[width=\linewidth]{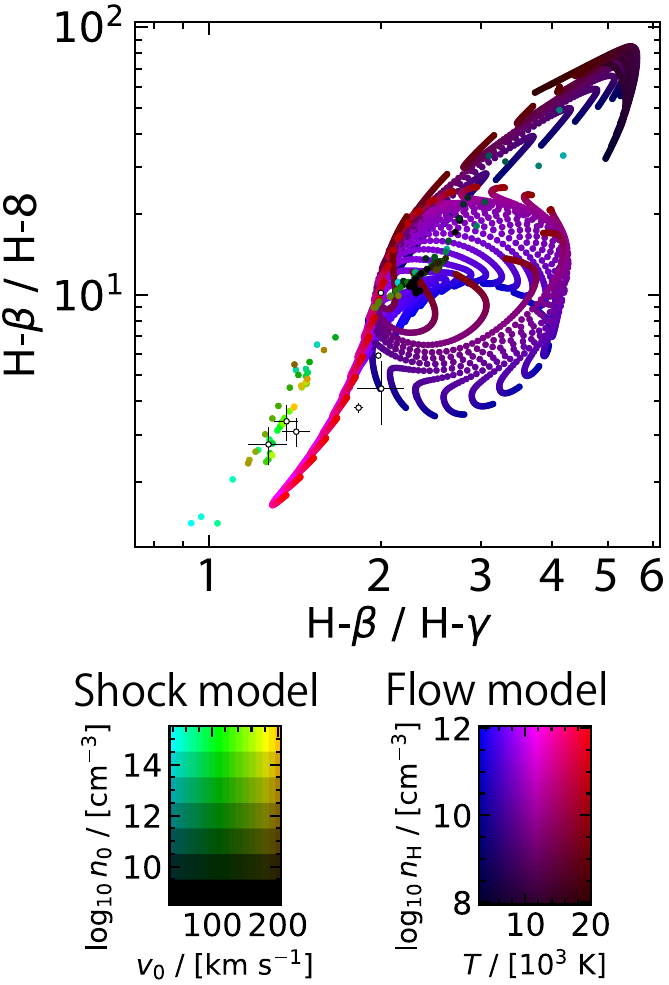} 
    \caption{
    Line ratios of objects with chromospheric dominated emission selected in Figure~\ref{fig:chr}.
    }\label{figA:chr}
\end{figure}

\section{Comments on 2M1115}\label{secA:2m1115}

 2MASS~J11151597$+$1937266 (hereafter 2M1115, spectral type: L2, $T_{\rm eff}$: 1700~K, log$(L_*/L_\sun)$: $-3.64$\footnote{The lunimosity is updated by using the GAIA DR3 distance.}, \citealp{Theissen2017-2m1115,Theissen2018-2m1115}; distance: 45~pc, \citealp{Gaia2016,Gaia2023}). The age is highly uncertain due to the fact that 2M1115 might not belong to any known young association or moving group. Given the presence of strong H$\alpha$ emissions and mid-infrared excess \citep{Theissen2017-2m1115,Theissen2018-2m1115} indicating the presence of surroundings, the age is assumed to be less than 50~Myr. As a result, the mass might be less than approximately 20~$M_{\rm Jup}$, estimated by the evolutionary model of \citet{Baraffe2003COND,Baraffe2015,Chabrier2023ATMO}. 

\begin{figure}[htbp]
    \includegraphics[width=\linewidth]{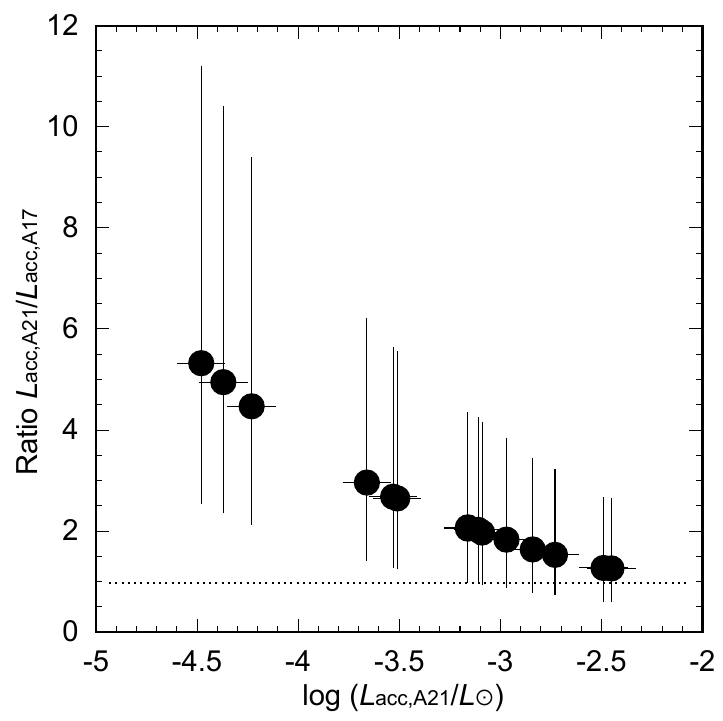} 
    \caption{
    \jhrev{Comparison of the accretion luminosity derived using the shock model ($L_{\rm acc,A21}$) with that obtained from the extrapolation of stellar scalings ($L_{\rm acc,A17}$) for the 15 data points located on the shock model loci in Figure~\ref{fig:plot-all}. It is the same as Figure \ref{fig:mdot}, but focuses on accretion luminosity.}
    }\label{figA:Lacc}
\end{figure}

\section{Line ratio of chromospheric dominated emission}\label{secA:chr}

Figure~\ref{figA:chr} shows line ratios of objects with chromospheric dominated emission selected in Figure~\ref{fig:chr}.

\jhrev{
\section{Ratio of accretion luminosity for the 15 objects exhibiting shock-dominated emission}\label{secA:Lacc}

Figure~\ref{figA:Lacc} shows the ratios of accretion luminosity, $L_{\rm acc,A21}$ and $L_{\rm acc,A17}$, derived from the shock model in \citet{Aoyama2021conversion} and from stellar scalings in \citet{alca2017}, respectively, for the 15 objects with shock-dominated emission selected in Figure~\ref{fig:plot-all}.
}

\section{Fraction of H$\beta$ luminosity to accretion luminosity with error}\label{secA:chr}

\begin{figure*}[htbp]
\centering
    \includegraphics[width=0.7\linewidth]{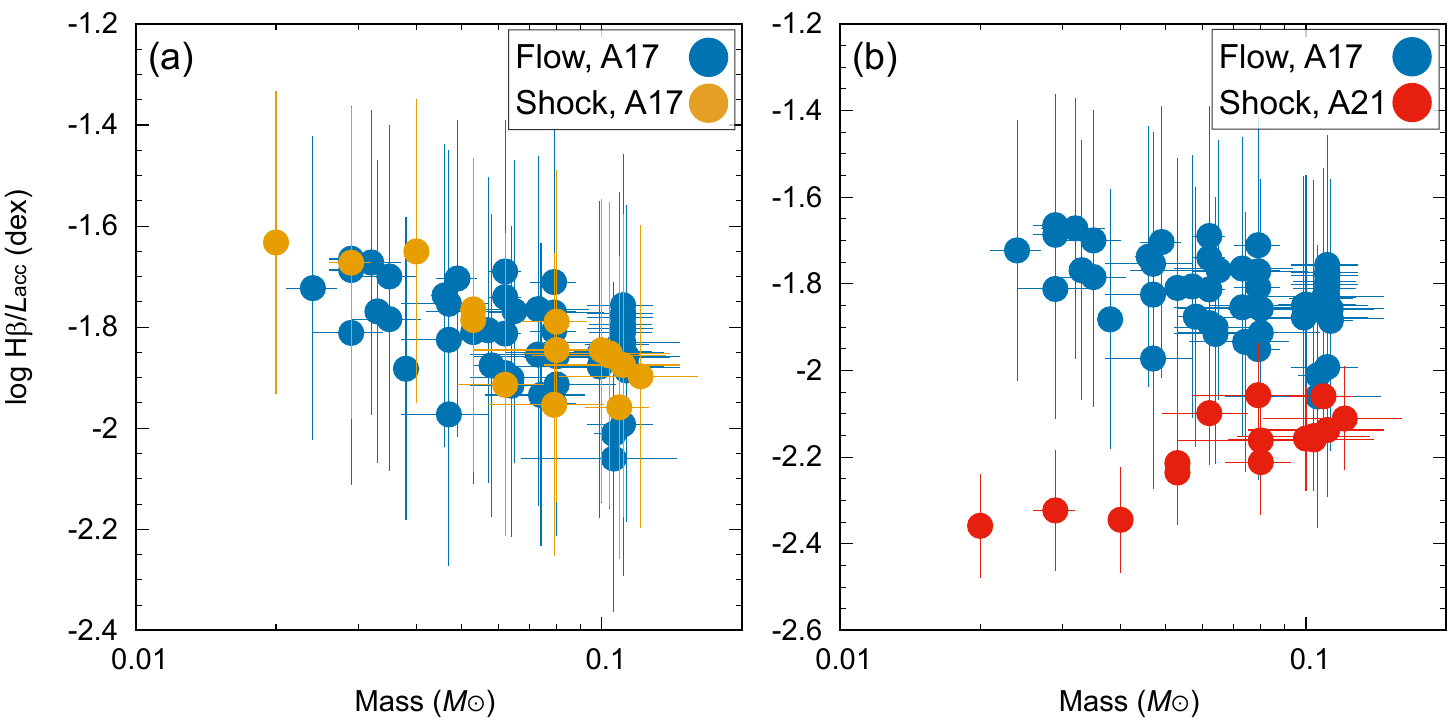} 
    \caption{
    Same figures with Figure~\ref{fig:hb_to_lacc}, but including the error.
    }\label{figA:hb-lacc_error}
\end{figure*}

Figure~\ref{figA:hb-lacc_error} shows same figures with Figure~\ref{fig:hb_to_lacc}, but including the error.

\section{Comparison in accretion luminosity and mass accretion rate}\label{secA:mdot}

\begin{figure*}[htbp]
\centering
    \includegraphics[width=0.7\linewidth]{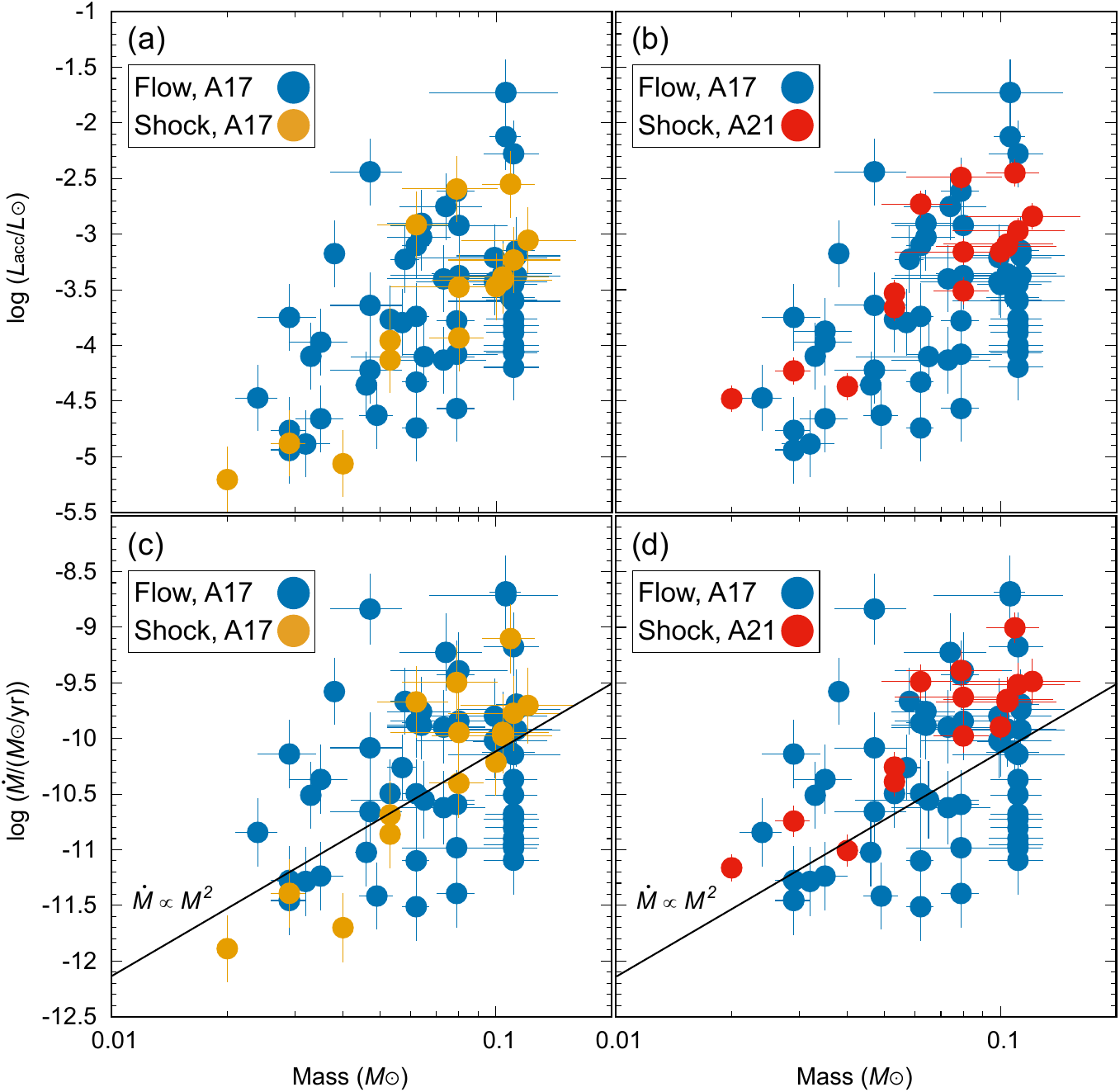} 
    \caption{
    Same with Figure~\ref{fig:param}, but only including accretion luminosity (upper panels) and mass accretion rate (bottom panels). The accretion luminosity and mass accretion rate of 15 data points, categorized as shock-dominated emissions in Figure~\ref{fig:plot-all}, are estimated by both the stellar scalings \citep{alca2017} (yellow) and the shock model \citep{Aoyama2021conversion} (red).
    }\label{figA:difference}
\end{figure*}

The accretion luminosity and mass accretion rate for 15 data points categorized as shock-dominated emissions in Figure~\ref{fig:plot-all} are estimated using two methods: the shock model \citep{Aoyama2021conversion} and stellar scalings \citep{alca2017}. Figure~\ref{figA:difference} illustrates the differences in their distributions as a function of mass. The estimation procedures for $L_{acc}$ and $\dot{M}$ are detailed in \S~\ref{subsec:mdot}. Notably, the values of $\dot{M}_{\rm A21}$ and $L_{\rm acc,A21}$ obtained from \citep{Aoyama2021conversion} are higher than those derived from the stellar scalings, $\dot{M}_{\rm A17}$ and $L_{\rm acc,A17}$ \citep{alca2017}, as discussed in \S~\ref{subsec:mdot}. In both cases, there is no clear boundary separating the shock and flow model categories (\S~\ref{subsec:boundary}).

\bibliography{sample63}{}

\begin{thebibliography}{}
\expandafter\ifx\csname natexlab\endcsname\relax\def\natexlab#1{#1}\fi
\providecommand{\url}[1]{\href{#1}{#1}}
\providecommand{\dodoi}[1]{doi:~\href{http://doi.org/#1}{\nolinkurl{#1}}}
\providecommand{\doeprint}[1]{\href{http://ascl.net/#1}{\nolinkurl{http://ascl.net/#1}}}
\providecommand{\doarXiv}[1]{\href{https://arxiv.org/abs/#1}{\nolinkurl{https://arxiv.org/abs/#1}}}

\bibitem[{{Alcal{\'a}} {et~al.}(2014){Alcal{\'a}}, {Natta}, {Manara}, {Spezzi},
  {Stelzer}, {Frasca}, {Biazzo}, {Covino}, {Randich}, {Rigliaco}, {Testi},
  {Comer{\'o}n}, {Cupani}, \& {D'Elia}}]{Alcala2014XSOOTER}
{Alcal{\'a}}, J.~M., {Natta}, A., {Manara}, C.~F., {et~al.} 2014, \aap, 561,
  A2, \dodoi{10.1051/0004-6361/201322254}

\bibitem[{{Alcal{\'a}} {et~al.}(2017){Alcal{\'a}}, {Manara}, {Natta}, {Frasca},
  {Testi}, {Nisini}, {Stelzer}, {Williams}, {Antoniucci}, {Biazzo}, {Covino},
  {Esposito}, {Getman}, \& {Rigliaco}}]{alca2017}
{Alcal{\'a}}, J.~M., {Manara}, C.~F., {Natta}, A., {et~al.} 2017, \aap, 600,
  A20, \dodoi{10.1051/0004-6361/201629929}

\bibitem[{{Almendros-Abad} {et~al.}(2024){Almendros-Abad}, {Manara}, {Testi},
  {Natta}, {Claes}, {Mu{\v{z}}i{\'c}}, {Sanchis}, {Alcal{\'a}}, {Bayo}, \&
  {Scholz}}]{Almendros-Abad2024YBD}
{Almendros-Abad}, V., {Manara}, C.~F., {Testi}, L., {et~al.} 2024, \aap, 685,
  A118, \dodoi{10.1051/0004-6361/202348649}

\bibitem[{{Aoyama} {et~al.}(2018){Aoyama}, {Ikoma}, \& {Tanigawa}}]{Aoyama2018}
{Aoyama}, Y., {Ikoma}, M., \& {Tanigawa}, T. 2018, \apj, 866, 84,
  \dodoi{10.3847/1538-4357/aadc11}

\bibitem[{{Aoyama} {et~al.}(2024){Aoyama}, {Marleau}, \&
  {Hashimoto}}]{Aoyama2024TWA27B}
{Aoyama}, Y., {Marleau}, G.-D., \& {Hashimoto}, J. 2024, \aj, 168, 155,
  \dodoi{10.3847/1538-3881/ad67df}

\bibitem[{{Aoyama} {et~al.}(2021){Aoyama}, {Marleau}, {Ikoma}, \&
  {Mordasini}}]{Aoyama2021conversion}
{Aoyama}, Y., {Marleau}, G.-D., {Ikoma}, M., \& {Mordasini}, C. 2021, \apjl,
  917, L30, \dodoi{10.3847/2041-8213/ac19bd}

\bibitem[{{Astropy Collaboration} {et~al.}(2013){Astropy Collaboration},
  {Robitaille}, {Tollerud}, {Greenfield}, {Droettboom}, {Bray}, {Aldcroft},
  {Davis}, {Ginsburg}, {Price-Whelan}, {Kerzendorf}, {Conley}, {Crighton},
  {Barbary}, {Muna}, {Ferguson}, {Grollier}, {Parikh}, {Nair}, {Unther},
  {Deil}, {Woillez}, {Conseil}, {Kramer}, {Turner}, {Singer}, {Fox}, {Weaver},
  {Zabalza}, {Edwards}, {Azalee Bostroem}, {Burke}, {Casey}, {Crawford},
  {Dencheva}, {Ely}, {Jenness}, {Labrie}, {Lim}, {Pierfederici}, {Pontzen},
  {Ptak}, {Refsdal}, {Servillat}, \& {Streicher}}]{Astropy2013}
{Astropy Collaboration}, {Robitaille}, T.~P., {Tollerud}, E.~J., {et~al.} 2013,
  \aap, 558, A33, \dodoi{10.1051/0004-6361/201322068}

\bibitem[{{Astropy Collaboration} {et~al.}(2018){Astropy Collaboration},
  {Price-Whelan}, {Sip{\H{o}}cz}, {G{\"u}nther}, {Lim}, {Crawford}, {Conseil},
  {Shupe}, {Craig}, {Dencheva}, {Ginsburg}, {VanderPlas}, {Bradley},
  {P{\'e}rez-Su{\'a}rez}, {de Val-Borro}, {Aldcroft}, {Cruz}, {Robitaille},
  {Tollerud}, {Ardelean}, {Babej}, {Bach}, {Bachetti}, {Bakanov}, {Bamford},
  {Barentsen}, {Barmby}, {Baumbach}, {Berry}, {Biscani}, {Boquien}, {Bostroem},
  {Bouma}, {Brammer}, {Bray}, {Breytenbach}, {Buddelmeijer}, {Burke},
  {Calderone}, {Cano Rodr{\'\i}guez}, {Cara}, {Cardoso}, {Cheedella}, {Copin},
  {Corrales}, {Crichton}, {D'Avella}, {Deil}, {Depagne}, {Dietrich}, {Donath},
  {Droettboom}, {Earl}, {Erben}, {Fabbro}, {Ferreira}, {Finethy}, {Fox},
  {Garrison}, {Gibbons}, {Goldstein}, {Gommers}, {Greco}, {Greenfield},
  {Groener}, {Grollier}, {Hagen}, {Hirst}, {Homeier}, {Horton}, {Hosseinzadeh},
  {Hu}, {Hunkeler}, {Ivezi{\'c}}, {Jain}, {Jenness}, {Kanarek}, {Kendrew},
  {Kern}, {Kerzendorf}, {Khvalko}, {King}, {Kirkby}, {Kulkarni}, {Kumar},
  {Lee}, {Lenz}, {Littlefair}, {Ma}, {Macleod}, {Mastropietro}, {McCully},
  {Montagnac}, {Morris}, {Mueller}, {Mumford}, {Muna}, {Murphy}, {Nelson},
  {Nguyen}, {Ninan}, {N{\"o}the}, {Ogaz}, {Oh}, {Parejko}, {Parley}, {Pascual},
  {Patil}, {Patil}, {Plunkett}, {Prochaska}, {Rastogi}, {Reddy Janga},
  {Sabater}, {Sakurikar}, {Seifert}, {Sherbert}, {Sherwood-Taylor}, {Shih},
  {Sick}, {Silbiger}, {Singanamalla}, {Singer}, {Sladen}, {Sooley},
  {Sornarajah}, {Streicher}, {Teuben}, {Thomas}, {Tremblay}, {Turner},
  {Terr{\'o}n}, {van Kerkwijk}, {de la Vega}, {Watkins}, {Weaver}, {Whitmore},
  {Woillez}, {Zabalza}, \& {Astropy Contributors}}]{Astropy2018}
{Astropy Collaboration}, {Price-Whelan}, A.~M., {Sip{\H{o}}cz}, B.~M., {et~al.}
  2018, \aj, 156, 123, \dodoi{10.3847/1538-3881/aabc4f}

\bibitem[{{Astropy Collaboration} {et~al.}(2022){Astropy Collaboration},
  {Price-Whelan}, {Lim}, {Earl}, {Starkman}, {Bradley}, {Shupe}, {Patil},
  {Corrales}, {Brasseur}, {N{\"o}the}, {Donath}, {Tollerud}, {Morris},
  {Ginsburg}, {Vaher}, {Weaver}, {Tocknell}, {Jamieson}, {van Kerkwijk},
  {Robitaille}, {Merry}, {Bachetti}, {G{\"u}nther}, {Aldcroft},
  {Alvarado-Montes}, {Archibald}, {B{\'o}di}, {Bapat}, {Barentsen},
  {Baz{\'a}n}, {Biswas}, {Boquien}, {Burke}, {Cara}, {Cara}, {Conroy},
  {Conseil}, {Craig}, {Cross}, {Cruz}, {D'Eugenio}, {Dencheva}, {Devillepoix},
  {Dietrich}, {Eigenbrot}, {Erben}, {Ferreira}, {Foreman-Mackey}, {Fox},
  {Freij}, {Garg}, {Geda}, {Glattly}, {Gondhalekar}, {Gordon}, {Grant},
  {Greenfield}, {Groener}, {Guest}, {Gurovich}, {Handberg}, {Hart},
  {Hatfield-Dodds}, {Homeier}, {Hosseinzadeh}, {Jenness}, {Jones}, {Joseph},
  {Kalmbach}, {Karamehmetoglu}, {Ka{\l}uszy{\'n}ski}, {Kelley}, {Kern},
  {Kerzendorf}, {Koch}, {Kulumani}, {Lee}, {Ly}, {Ma}, {MacBride}, {Maljaars},
  {Muna}, {Murphy}, {Norman}, {O'Steen}, {Oman}, {Pacifici}, {Pascual},
  {Pascual-Granado}, {Patil}, {Perren}, {Pickering}, {Rastogi}, {Roulston},
  {Ryan}, {Rykoff}, {Sabater}, {Sakurikar}, {Salgado}, {Sanghi}, {Saunders},
  {Savchenko}, {Schwardt}, {Seifert-Eckert}, {Shih}, {Jain}, {Shukla}, {Sick},
  {Simpson}, {Singanamalla}, {Singer}, {Singhal}, {Sinha}, {Sip{\H{o}}cz},
  {Spitler}, {Stansby}, {Streicher}, {{\v{S}}umak}, {Swinbank}, {Taranu},
  {Tewary}, {Tremblay}, {de Val-Borro}, {Van Kooten}, {Vasovi{\'c}}, {Verma},
  {de Miranda Cardoso}, {Williams}, {Wilson}, {Winkel}, {Wood-Vasey}, {Xue},
  {Yoachim}, {Zhang}, {Zonca}, \& {Astropy Project Contributors}}]{Astropy2022}
{Astropy Collaboration}, {Price-Whelan}, A.~M., {Lim}, P.~L., {et~al.} 2022,
  \apj, 935, 167, \dodoi{10.3847/1538-4357/ac7c74}

\bibitem[{{Baraffe} {et~al.}(2003){Baraffe}, {Chabrier}, {Barman}, {Allard}, \&
  {Hauschildt}}]{Baraffe2003COND}
{Baraffe}, I., {Chabrier}, G., {Barman}, T.~S., {Allard}, F., \& {Hauschildt},
  P.~H. 2003, \aap, 402, 701, \dodoi{10.1051/0004-6361:20030252}

\bibitem[{{Baraffe} {et~al.}(2015){Baraffe}, {Homeier}, {Allard}, \&
  {Chabrier}}]{Baraffe2015}
{Baraffe}, I., {Homeier}, D., {Allard}, F., \& {Chabrier}, G. 2015, \aap, 577,
  A42, \dodoi{10.1051/0004-6361/201425481}

\bibitem[{{Barrado y Navascu{\'e}s} \& {Mart{\'\i}n}(2003)}]{Barrado2003EW}
{Barrado y Navascu{\'e}s}, D., \& {Mart{\'\i}n}, E.~L. 2003, \aj, 126, 2997,
  \dodoi{10.1086/379673}

\bibitem[{{Bertout} {et~al.}(1982){Bertout}, {Wolf}, {Carrasco}, \&
  {Mundt}}]{Bertout1982Accretion}
{Bertout}, C., {Wolf}, B., {Carrasco}, L., \& {Mundt}, R. 1982, \aaps, 47, 419

\bibitem[{{Betti} {et~al.}(2022){Betti}, {Follette}, {Ward-Duong}, {Aoyama},
  {Marleau}, {Bary}, {Robinson}, {Janson}, {Balmer}, {Chauvin}, \&
  {Palma-Bifani}}]{Betti2022D1b}
{Betti}, S.~K., {Follette}, K.~B., {Ward-Duong}, K., {et~al.} 2022, \apjl, 935,
  L18, \dodoi{10.3847/2041-8213/ac85ef}

\bibitem[{{Betti} {et~al.}(2023){Betti}, {Follette}, {Ward-Duong}, {Peck},
  {Aoyama}, {Bary}, {Dacus}, {Edwards}, {Marleau}, {Mohamed}, {Palmo},
  {Plunkett}, {Robinson}, \& {Wang}}]{Betti2023Survey}
---. 2023, \aj, 166, 262, \dodoi{10.3847/1538-3881/ad06b8}

\bibitem[{{Calvet} \& {Gullbring}(1998)}]{Calvet1998ff}
{Calvet}, N., \& {Gullbring}, E. 1998, The Astrophysical Journal, 509, 802,
  \dodoi{10.1086/306527}

\bibitem[{{Calvet} \& {Hartmann}(1992)}]{Calvet1992Model}
{Calvet}, N., \& {Hartmann}, L. 1992, \apj, 386, 239, \dodoi{10.1086/171010}

\bibitem[{{Cardelli} {et~al.}(1989){Cardelli}, {Clayton}, \&
  {Mathis}}]{Cardelli1989Extinction}
{Cardelli}, J.~A., {Clayton}, G.~C., \& {Mathis}, J.~S. 1989, \apj, 345, 245,
  \dodoi{10.1086/167900}

\bibitem[{{Chabrier} {et~al.}(2023){Chabrier}, {Baraffe}, {Phillips}, \&
  {Debras}}]{Chabrier2023ATMO}
{Chabrier}, G., {Baraffe}, I., {Phillips}, M., \& {Debras}, F. 2023, \aap, 671,
  A119, \dodoi{10.1051/0004-6361/202243832}

\bibitem[{{Chauvin} {et~al.}(2004){Chauvin}, {Lagrange}, {Dumas}, {Zuckerman},
  {Mouillet}, {Song}, {Beuzit}, \& {Lowrance}}]{Chauvin2004}
{Chauvin}, G., {Lagrange}, A.~M., {Dumas}, C., {et~al.} 2004, \aap, 425, L29,
  \dodoi{10.1051/0004-6361:200400056}

\bibitem[{{Delorme} {et~al.}(2013){Delorme}, {Gagn{\'e}}, {Girard}, {Lagrange},
  {Chauvin}, {Naud}, {Lafreni{\`e}re}, {Doyon}, {Riedel}, {Bonnefoy}, \&
  {Malo}}]{Delorme2013D1b}
{Delorme}, P., {Gagn{\'e}}, J., {Girard}, J.~H., {et~al.} 2013, \aap, 553, L5,
  \dodoi{10.1051/0004-6361/201321169}

\bibitem[{{Demars} {et~al.}(2023){Demars}, {Bonnefoy}, {Dougados}, {Aoyama},
  {Thanathibodee}, {Marleau}, {Tremblin}, {Delorme}, {Palma-Bifani}, {Petrus},
  {Bowler}, {Chauvin}, \& {Lagrange}}]{Demars2023Variability}
{Demars}, D., {Bonnefoy}, M., {Dougados}, C., {et~al.} 2023, \aap, 676, A123,
  \dodoi{10.1051/0004-6361/202346221}

\bibitem[{{Edwards} {et~al.}(1994){Edwards}, {Hartigan}, {Ghandour}, \&
  {Andrulis}}]{Edwards1994Accretion}
{Edwards}, S., {Hartigan}, P., {Ghandour}, L., \& {Andrulis}, C. 1994, \aj,
  108, 1056, \dodoi{10.1086/117134}

\bibitem[{{Fang} {et~al.}(2023){Fang}, {Pascucci}, {Edwards}, {Gorti},
  {Hillenbrand}, \& {Carpenter}}]{Fang2023USco}
{Fang}, M., {Pascucci}, I., {Edwards}, S., {et~al.} 2023, \apj, 945, 112,
  \dodoi{10.3847/1538-4357/acb2c9}

\bibitem[{{Fischer} {et~al.}(2023){Fischer}, {Hillenbrand}, {Herczeg},
  {Johnstone}, {Kospal}, \& {Dunham}}]{Fischer2023PPVII_variability}
{Fischer}, W.~J., {Hillenbrand}, L.~A., {Herczeg}, G.~J., {et~al.} 2023, in
  Astronomical Society of the Pacific Conference Series, Vol. 534, Protostars
  and Planets VII, ed. S.~{Inutsuka}, Y.~{Aikawa}, T.~{Muto}, K.~{Tomida}, \&
  M.~{Tamura}, 355, \dodoi{10.48550/arXiv.2203.11257}

\bibitem[{{Gaia Collaboration} {et~al.}(2016){Gaia Collaboration}, {Prusti},
  {de Bruijne}, {Brown}, {Vallenari}, {Babusiaux}, {Bailer-Jones}, {Bastian},
  {Biermann}, {Evans}, {Eyer}, {Jansen}, {Jordi}, {Klioner}, {Lammers},
  {Lindegren}, {Luri}, {Mignard}, {Milligan}, {Panem}, {Poinsignon},
  {Pourbaix}, {Randich}, {Sarri}, {Sartoretti}, {Siddiqui}, {Soubiran},
  {Valette}, {van Leeuwen}, {Walton}, {Aerts}, {Arenou}, {Cropper}, {Drimmel},
  {H{\o}g}, {Katz}, {Lattanzi}, {O'Mullane}, {Grebel}, {Holland}, {Huc},
  {Passot}, {Bramante}, {Cacciari}, {Casta{\~n}eda}, {Chaoul}, {Cheek}, {De
  Angeli}, {Fabricius}, {Guerra}, {Hern{\'a}ndez}, {Jean-Antoine-Piccolo},
  {Masana}, {Messineo}, {Mowlavi}, {Nienartowicz}, {Ord{\'o}{\~n}ez-Blanco},
  {Panuzzo}, {Portell}, {Richards}, {Riello}, {Seabroke}, {Tanga},
  {Th{\'e}venin}, {Torra}, {Els}, {Gracia-Abril}, {Comoretto},
  {Garcia-Reinaldos}, {Lock}, {Mercier}, {Altmann}, {Andrae}, {Astraatmadja},
  {Bellas-Velidis}, {Benson}, {Berthier}, {Blomme}, {Busso}, {Carry},
  {Cellino}, {Clementini}, {Cowell}, {Creevey}, {Cuypers}, {Davidson}, {De
  Ridder}, {de Torres}, {Delchambre}, {Dell'Oro}, {Ducourant}, {Fr{\'e}mat},
  {Garc{\'\i}a-Torres}, {Gosset}, {Halbwachs}, {Hambly}, {Harrison}, {Hauser},
  {Hestroffer}, {Hodgkin}, {Huckle}, {Hutton}, {Jasniewicz}, {Jordan},
  {Kontizas}, {Korn}, {Lanzafame}, {Manteiga}, {Moitinho}, {Muinonen},
  {Osinde}, {Pancino}, {Pauwels}, {Petit}, {Recio-Blanco}, {Robin}, {Sarro},
  {Siopis}, {Smith}, {Smith}, {Sozzetti}, {Thuillot}, {van Reeven}, {Viala},
  {Abbas}, {Abreu Aramburu}, {Accart}, {Aguado}, {Allan}, {Allasia},
  {Altavilla}, {{\'A}lvarez}, {Alves}, {Anderson}, {Andrei}, {Anglada Varela},
  {Antiche}, {Antoja}, {Ant{\'o}n}, {Arcay}, {Atzei}, {Ayache}, {Bach},
  {Baker}, {Balaguer-N{\'u}{\~n}ez}, {Barache}, {Barata}, {Barbier}, {Barblan},
  {Baroni}, {Barrado y Navascu{\'e}s}, {Barros}, {Barstow}, {Becciani},
  {Bellazzini}, {Bellei}, {Bello Garc{\'\i}a}, {Belokurov}, {Bendjoya},
  {Berihuete}, {Bianchi}, {Bienaym{\'e}}, {Billebaud}, {Blagorodnova},
  {Blanco-Cuaresma}, {Boch}, {Bombrun}, {Borrachero}, {Bouquillon}, {Bourda},
  {Bouy}, {Bragaglia}, {Breddels}, {Brouillet}, {Br{\"u}semeister},
  {Bucciarelli}, {Budnik}, {Burgess}, {Burgon}, {Burlacu}, {Busonero}, {Buzzi},
  {Caffau}, {Cambras}, {Campbell}, {Cancelliere}, {Cantat-Gaudin}, {Carlucci},
  {Carrasco}, {Castellani}, {Charlot}, {Charnas}, {Charvet}, {Chassat},
  {Chiavassa}, {Clotet}, {Cocozza}, {Collins}, {Collins}, {Costigan}, {Crifo},
  {Cross}, {Crosta}, {Crowley}, {Dafonte}, {Damerdji}, {Dapergolas}, {David},
  {David}, {De Cat}, {de Felice}, {de Laverny}, {De Luise}, {De March}, {de
  Martino}, {de Souza}, {Debosscher}, {del Pozo}, {Delbo}, {Delgado},
  {Delgado}, {di Marco}, {Di Matteo}, {Diakite}, {Distefano}, {Dolding}, {Dos
  Anjos}, {Drazinos}, {Dur{\'a}n}, {Dzigan}, {Ecale}, {Edvardsson}, {Enke},
  {Erdmann}, {Escolar}, {Espina}, {Evans}, {Eynard Bontemps}, {Fabre},
  {Fabrizio}, {Faigler}, {Falc{\~a}o}, {Farr{\`a}s Casas}, {Faye}, {Federici},
  {Fedorets}, {Fern{\'a}ndez-Hern{\'a}ndez}, {Fernique}, {Fienga}, {Figueras},
  {Filippi}, {Findeisen}, {Fonti}, {Fouesneau}, {Fraile}, {Fraser}, {Fuchs},
  {Furnell}, {Gai}, {Galleti}, {Galluccio}, {Garabato}, {Garc{\'\i}a-Sedano},
  {Gar{\'e}}, {Garofalo}, {Garralda}, {Gavras}, {Gerssen}, {Geyer}, {Gilmore},
  {Girona}, {Giuffrida}, {Gomes}, {Gonz{\'a}lez-Marcos},
  {Gonz{\'a}lez-N{\'u}{\~n}ez}, {Gonz{\'a}lez-Vidal}, {Granvik}, {Guerrier},
  {Guillout}, {Guiraud}, {G{\'u}rpide}, {Guti{\'e}rrez-S{\'a}nchez}, {Guy},
  {Haigron}, {Hatzidimitriou}, {Haywood}, {Heiter}, {Helmi}, {Hobbs},
  {Hofmann}, {Holl}, {Holland}, {Hunt}, {Hypki}, {Icardi}, {Irwin}, {Jevardat
  de Fombelle}, {Jofr{\'e}}, {Jonker}, {Jorissen}, {Julbe}, {Karampelas},
  {Kochoska}, {Kohley}, {Kolenberg}, {Kontizas}, {Koposov}, {Kordopatis},
  {Koubsky}, {Kowalczyk}, {Krone-Martins}, {Kudryashova}, {Kull}, {Bachchan},
  {Lacoste-Seris}, {Lanza}, {Lavigne}, {Le Poncin-Lafitte}, {Lebreton},
  {Lebzelter}, {Leccia}, {Leclerc}, {Lecoeur-Taibi}, {Lemaitre}, {Lenhardt},
  {Leroux}, {Liao}, {Licata}, {Lindstr{\o}m}, {Lister}, {Livanou}, {Lobel},
  {L{\"o}ffler}, {L{\'o}pez}, {Lopez-Lozano}, {Lorenz}, {Loureiro},
  {MacDonald}, {Magalh{\~a}es Fernandes}, {Managau}, {Mann}, {Mantelet},
  {Marchal}, {Marchant}, {Marconi}, {Marie}, {Marinoni}, {Marrese},
  {Marschalk{\'o}}, {Marshall}, {Mart{\'\i}n-Fleitas}, {Martino}, {Mary},
  {Matijevi{\v{c}}}, {Mazeh}, {McMillan}, {Messina}, {Mestre}, {Michalik},
  {Millar}, {Miranda}, {Molina}, {Molinaro}, {Molinaro}, {Moln{\'a}r},
  {Moniez}, {Montegriffo}, {Monteiro}, {Mor}, {Mora}, {Morbidelli}, {Morel},
  {Morgenthaler}, {Morley}, {Morris}, {Mulone}, {Muraveva}, {Musella},
  {Narbonne}, {Nelemans}, {Nicastro}, {Noval}, {Ord{\'e}novic},
  {Ordieres-Mer{\'e}}, {Osborne}, {Pagani}, {Pagano}, {Pailler}, {Palacin},
  {Palaversa}, {Parsons}, {Paulsen}, {Pecoraro}, {Pedrosa}, {Pentik{\"a}inen},
  {Pereira}, {Pichon}, {Piersimoni}, {Pineau}, {Plachy}, {Plum}, {Poujoulet},
  {Pr{\v{s}}a}, {Pulone}, {Ragaini}, {Rago}, {Rambaux}, {Ramos-Lerate},
  {Ranalli}, {Rauw}, {Read}, {Regibo}, {Renk}, {Reyl{\'e}}, {Ribeiro},
  {Rimoldini}, {Ripepi}, {Riva}, {Rixon}, {Roelens}, {Romero-G{\'o}mez},
  {Rowell}, {Royer}, {Rudolph}, {Ruiz-Dern}, {Sadowski}, {Sagrist{\`a}
  Sell{\'e}s}, {Sahlmann}, {Salgado}, {Salguero}, {Sarasso}, {Savietto},
  {Schnorhk}, {Schultheis}, {Sciacca}, {Segol}, {Segovia}, {Segransan},
  {Serpell}, {Shih}, {Smareglia}, {Smart}, {Smith}, {Solano}, {Solitro},
  {Sordo}, {Soria Nieto}, {Souchay}, {Spagna}, {Spoto}, {Stampa}, {Steele},
  {Steidelm{\"u}ller}, {Stephenson}, {Stoev}, {Suess}, {S{\"u}veges}, {Surdej},
  {Szabados}, {Szegedi-Elek}, {Tapiador}, {Taris}, {Tauran}, {Taylor},
  {Teixeira}, {Terrett}, {Tingley}, {Trager}, {Turon}, {Ulla}, {Utrilla},
  {Valentini}, {van Elteren}, {Van Hemelryck}, {van Leeuwen}, {Varadi},
  {Vecchiato}, {Veljanoski}, {Via}, {Vicente}, {Vogt}, {Voss}, {Votruba},
  {Voutsinas}, {Walmsley}, {Weiler}, {Weingrill}, {Werner}, {Wevers},
  {Whitehead}, {Wyrzykowski}, {Yoldas}, {{\v{Z}}erjal}, {Zucker}, {Zurbach},
  {Zwitter}, {Alecu}, {Allen}, {Allende Prieto}, {Amorim},
  {Anglada-Escud{\'e}}, {Arsenijevic}, {Azaz}, {Balm}, {Beck}, {Bernstein},
  {Bigot}, {Bijaoui}, {Blasco}, {Bonfigli}, {Bono}, {Boudreault}, {Bressan},
  {Brown}, {Brunet}, {Bunclark}, {Buonanno}, {Butkevich}, {Carret}, {Carrion},
  {Chemin}, {Ch{\'e}reau}, {Corcione}, {Darmigny}, {de Boer}, {de Teodoro}, {de
  Zeeuw}, {Delle Luche}, {Domingues}, {Dubath}, {Fodor}, {Fr{\'e}zouls},
  {Fries}, {Fustes}, {Fyfe}, {Gallardo}, {Gallegos}, {Gardiol}, {Gebran},
  {Gomboc}, {G{\'o}mez}, {Grux}, {Gueguen}, {Heyrovsky}, {Hoar}, {Iannicola},
  {Isasi Parache}, {Janotto}, {Joliet}, {Jonckheere}, {Keil}, {Kim},
  {Klagyivik}, {Klar}, {Knude}, {Kochukhov}, {Kolka}, {Kos}, {Kutka}, {Lainey},
  {LeBouquin}, {Liu}, {Loreggia}, {Makarov}, {Marseille}, {Martayan},
  {Martinez-Rubi}, {Massart}, {Meynadier}, {Mignot}, {Munari}, {Nguyen},
  {Nordlander}, {Ocvirk}, {O'Flaherty}, {Olias Sanz}, {Ortiz}, {Osorio},
  {Oszkiewicz}, {Ouzounis}, {Palmer}, {Park}, {Pasquato}, {Peltzer}, {Peralta},
  {P{\'e}turaud}, {Pieniluoma}, {Pigozzi}, {Poels}, {Prat}, {Prod'homme},
  {Raison}, {Rebordao}, {Risquez}, {Rocca-Volmerange}, {Rosen}, {Ruiz-Fuertes},
  {Russo}, {Sembay}, {Serraller Vizcaino}, {Short}, {Siebert}, {Silva},
  {Sinachopoulos}, {Slezak}, {Soffel}, {Sosnowska}, {Strai{\v{z}}ys}, {ter
  Linden}, {Terrell}, {Theil}, {Tiede}, {Troisi}, {Tsalmantza}, {Tur},
  {Vaccari}, {Vachier}, {Valles}, {Van Hamme}, {Veltz}, {Virtanen}, {Wallut},
  {Wichmann}, {Wilkinson}, {Ziaeepour}, \& {Zschocke}}]{Gaia2016}
{Gaia Collaboration}, {Prusti}, T., {de Bruijne}, J.~H.~J., {et~al.} 2016,
  \aap, 595, A1, \dodoi{10.1051/0004-6361/201629272}

\bibitem[{{Gaia Collaboration} {et~al.}(2023){Gaia Collaboration}, {Vallenari},
  {Brown}, {Prusti}, {de Bruijne}, {Arenou}, {Babusiaux}, {Biermann},
  {Creevey}, {Ducourant}, {Evans}, {Eyer}, {Guerra}, {Hutton}, {Jordi},
  {Klioner}, {Lammers}, {Lindegren}, {Luri}, {Mignard}, {Panem}, {Pourbaix},
  {Randich}, {Sartoretti}, {Soubiran}, {Tanga}, {Walton}, {Bailer-Jones},
  {Bastian}, {Drimmel}, {Jansen}, {Katz}, {Lattanzi}, {van Leeuwen}, {Bakker},
  {Cacciari}, {Casta{\~n}eda}, {De Angeli}, {Fabricius}, {Fouesneau},
  {Fr{\'e}mat}, {Galluccio}, {Guerrier}, {Heiter}, {Masana}, {Messineo},
  {Mowlavi}, {Nicolas}, {Nienartowicz}, {Pailler}, {Panuzzo}, {Riclet}, {Roux},
  {Seabroke}, {Sordo}, {Th{\'e}venin}, {Gracia-Abril}, {Portell}, {Teyssier},
  {Altmann}, {Andrae}, {Audard}, {Bellas-Velidis}, {Benson}, {Berthier},
  {Blomme}, {Burgess}, {Busonero}, {Busso}, {C{\'a}novas}, {Carry}, {Cellino},
  {Cheek}, {Clementini}, {Damerdji}, {Davidson}, {de Teodoro}, {Nu{\~n}ez
  Campos}, {Delchambre}, {Dell'Oro}, {Esquej}, {Fern{\'a}ndez-Hern{\'a}ndez},
  {Fraile}, {Garabato}, {Garc{\'\i}a-Lario}, {Gosset}, {Haigron}, {Halbwachs},
  {Hambly}, {Harrison}, {Hern{\'a}ndez}, {Hestroffer}, {Hodgkin}, {Holl},
  {Jan{\ss}en}, {Jevardat de Fombelle}, {Jordan}, {Krone-Martins}, {Lanzafame},
  {L{\"o}ffler}, {Marchal}, {Marrese}, {Moitinho}, {Muinonen}, {Osborne},
  {Pancino}, {Pauwels}, {Recio-Blanco}, {Reyl{\'e}}, {Riello}, {Rimoldini},
  {Roegiers}, {Rybizki}, {Sarro}, {Siopis}, {Smith}, {Sozzetti}, {Utrilla},
  {van Leeuwen}, {Abbas}, {{\'A}brah{\'a}m}, {Abreu Aramburu}, {Aerts},
  {Aguado}, {Ajaj}, {Aldea-Montero}, {Altavilla}, {{\'A}lvarez}, {Alves},
  {Anders}, {Anderson}, {Anglada Varela}, {Antoja}, {Baines}, {Baker},
  {Balaguer-N{\'u}{\~n}ez}, {Balbinot}, {Balog}, {Barache}, {Barbato},
  {Barros}, {Barstow}, {Bartolom{\'e}}, {Bassilana}, {Bauchet}, {Becciani},
  {Bellazzini}, {Berihuete}, {Bernet}, {Bertone}, {Bianchi}, {Binnenfeld},
  {Blanco-Cuaresma}, {Blazere}, {Boch}, {Bombrun}, {Bossini}, {Bouquillon},
  {Bragaglia}, {Bramante}, {Breedt}, {Bressan}, {Brouillet}, {Brugaletta},
  {Bucciarelli}, {Burlacu}, {Butkevich}, {Buzzi}, {Caffau}, {Cancelliere},
  {Cantat-Gaudin}, {Carballo}, {Carlucci}, {Carnerero}, {Carrasco},
  {Casamiquela}, {Castellani}, {Castro-Ginard}, {Chaoul}, {Charlot}, {Chemin},
  {Chiaramida}, {Chiavassa}, {Chornay}, {Comoretto}, {Contursi}, {Cooper},
  {Cornez}, {Cowell}, {Crifo}, {Cropper}, {Crosta}, {Crowley}, {Dafonte},
  {Dapergolas}, {David}, {David}, {de Laverny}, {De Luise}, {De March}, {De
  Ridder}, {de Souza}, {de Torres}, {del Peloso}, {del Pozo}, {Delbo},
  {Delgado}, {Delisle}, {Demouchy}, {Dharmawardena}, {Di Matteo}, {Diakite},
  {Diener}, {Distefano}, {Dolding}, {Edvardsson}, {Enke}, {Fabre}, {Fabrizio},
  {Faigler}, {Fedorets}, {Fernique}, {Fienga}, {Figueras}, {Fournier},
  {Fouron}, {Fragkoudi}, {Gai}, {Garcia-Gutierrez}, {Garcia-Reinaldos},
  {Garc{\'\i}a-Torres}, {Garofalo}, {Gavel}, {Gavras}, {Gerlach}, {Geyer},
  {Giacobbe}, {Gilmore}, {Girona}, {Giuffrida}, {Gomel}, {Gomez},
  {Gonz{\'a}lez-N{\'u}{\~n}ez}, {Gonz{\'a}lez-Santamar{\'\i}a},
  {Gonz{\'a}lez-Vidal}, {Granvik}, {Guillout}, {Guiraud},
  {Guti{\'e}rrez-S{\'a}nchez}, {Guy}, {Hatzidimitriou}, {Hauser}, {Haywood},
  {Helmer}, {Helmi}, {Sarmiento}, {Hidalgo}, {Hilger}, {H{\l}adczuk}, {Hobbs},
  {Holland}, {Huckle}, {Jardine}, {Jasniewicz}, {Jean-Antoine Piccolo},
  {Jim{\'e}nez-Arranz}, {Jorissen}, {Juaristi Campillo}, {Julbe}, {Karbevska},
  {Kervella}, {Khanna}, {Kontizas}, {Kordopatis}, {Korn}, {K{\'o}sp{\'a}l},
  {Kostrzewa-Rutkowska}, {Kruszy{\'n}ska}, {Kun}, {Laizeau}, {Lambert},
  {Lanza}, {Lasne}, {Le Campion}, {Lebreton}, {Lebzelter}, {Leccia}, {Leclerc},
  {Lecoeur-Taibi}, {Liao}, {Licata}, {Lindstr{\o}m}, {Lister}, {Livanou},
  {Lobel}, {Lorca}, {Loup}, {Madrero Pardo}, {Magdaleno Romeo}, {Managau},
  {Mann}, {Manteiga}, {Marchant}, {Marconi}, {Marcos}, {Marcos Santos},
  {Mar{\'\i}n Pina}, {Marinoni}, {Marocco}, {Marshall}, {Martin Polo},
  {Mart{\'\i}n-Fleitas}, {Marton}, {Mary}, {Masip}, {Massari},
  {Mastrobuono-Battisti}, {Mazeh}, {McMillan}, {Messina}, {Michalik}, {Millar},
  {Mints}, {Molina}, {Molinaro}, {Moln{\'a}r}, {Monari}, {Mongui{\'o}},
  {Montegriffo}, {Montero}, {Mor}, {Mora}, {Morbidelli}, {Morel}, {Morris},
  {Muraveva}, {Murphy}, {Musella}, {Nagy}, {Noval}, {Oca{\~n}a}, {Ogden},
  {Ordenovic}, {Osinde}, {Pagani}, {Pagano}, {Palaversa}, {Palicio},
  {Pallas-Quintela}, {Panahi}, {Payne-Wardenaar}, {Pe{\~n}alosa Esteller},
  {Penttil{\"a}}, {Pichon}, {Piersimoni}, {Pineau}, {Plachy}, {Plum}, {Poggio},
  {Pr{\v{s}}a}, {Pulone}, {Racero}, {Ragaini}, {Rainer}, {Raiteri}, {Rambaux},
  {Ramos}, {Ramos-Lerate}, {Re Fiorentin}, {Regibo}, {Richards}, {Rios Diaz},
  {Ripepi}, {Riva}, {Rix}, {Rixon}, {Robichon}, {Robin}, {Robin}, {Roelens},
  {Rogues}, {Rohrbasser}, {Romero-G{\'o}mez}, {Rowell}, {Royer}, {Ruz Mieres},
  {Rybicki}, {Sadowski}, {S{\'a}ez N{\'u}{\~n}ez}, {Sagrist{\`a} Sell{\'e}s},
  {Sahlmann}, {Salguero}, {Samaras}, {Sanchez Gimenez}, {Sanna},
  {Santove{\~n}a}, {Sarasso}, {Schultheis}, {Sciacca}, {Segol}, {Segovia},
  {S{\'e}gransan}, {Semeux}, {Shahaf}, {Siddiqui}, {Siebert}, {Siltala},
  {Silvelo}, {Slezak}, {Slezak}, {Smart}, {Snaith}, {Solano}, {Solitro},
  {Souami}, {Souchay}, {Spagna}, {Spina}, {Spoto}, {Steele},
  {Steidelm{\"u}ller}, {Stephenson}, {S{\"u}veges}, {Surdej}, {Szabados},
  {Szegedi-Elek}, {Taris}, {Taylor}, {Teixeira}, {Tolomei}, {Tonello}, {Torra},
  {Torra}, {Torralba Elipe}, {Trabucchi}, {Tsounis}, {Turon}, {Ulla}, {Unger},
  {Vaillant}, {van Dillen}, {van Reeven}, {Vanel}, {Vecchiato}, {Viala},
  {Vicente}, {Voutsinas}, {Weiler}, {Wevers}, {Wyrzykowski}, {Yoldas}, {Yvard},
  {Zhao}, {Zorec}, {Zucker}, \& {Zwitter}}]{Gaia2023}
{Gaia Collaboration}, {Vallenari}, A., {Brown}, A.~G.~A., {et~al.} 2023, \aap,
  674, A1, \dodoi{10.1051/0004-6361/202243940}

\bibitem[{{Gullbring} {et~al.}(1998){Gullbring}, {Hartmann}, {Brice{\~n}o}, \&
  {Calvet}}]{Gullbring1998TTS}
{Gullbring}, E., {Hartmann}, L., {Brice{\~n}o}, C., \& {Calvet}, N. 1998, \apj,
  492, 323, \dodoi{10.1086/305032}

\bibitem[{{Hartmann} {et~al.}(2016){Hartmann}, {Herczeg}, \&
  {Calvet}}]{Hartmann2016accretion}
{Hartmann}, L., {Herczeg}, G., \& {Calvet}, N. 2016, \araa, 54, 135,
  \dodoi{10.1146/annurev-astro-081915-023347}

\bibitem[{{Hartmann} {et~al.}(1994){Hartmann}, {Hewett}, \&
  {Calvet}}]{Hartmann1994Model}
{Hartmann}, L., {Hewett}, R., \& {Calvet}, N. 1994, \apj, 426, 669,
  \dodoi{10.1086/174104}

\bibitem[{{Herczeg} \& {Hillenbrand}(2008)}]{Herczeg2008ff}
{Herczeg}, G.~J., \& {Hillenbrand}, L.~A. 2008, \apj, 681, 594,
  \dodoi{10.1086/586728}

\bibitem[{Hunter(2007)}]{Hunter2007Matplotlib}
Hunter, J.~D. 2007, Computing in Science \& Engineering, 9, 90,
  \dodoi{10.1109/MCSE.2007.55}

\bibitem[{{Joy}(1945)}]{Joy1945TTS}
{Joy}, A.~H. 1945, \apj, 102, 168, \dodoi{10.1086/144749}

\bibitem[{{Kwan} \& {Fischer}(2011)}]{Kwan2011Flow}
{Kwan}, J., \& {Fischer}, W. 2011, \mnras, 411, 2383,
  \dodoi{10.1111/j.1365-2966.2010.17863.x}

\bibitem[{{Luhman} {et~al.}(2023){Luhman}, {Tremblin}, {Birkmann},
  {Manjavacas}, {Valenti}, {Alves de Oliveira}, {Beck}, {Giardino},
  {L{\"u}tzgendorf}, {Rauscher}, \& {Sirianni}}]{Luhman2023TWA27b}
{Luhman}, K.~L., {Tremblin}, P., {Birkmann}, S.~M., {et~al.} 2023, \apjl, 949,
  L36, \dodoi{10.3847/2041-8213/acd635}

\bibitem[{{Manara} {et~al.}(2017){Manara}, {Frasca}, {Alcal{\'a}}, {Natta},
  {Stelzer}, \& {Testi}}]{Manara2017Chr}
{Manara}, C.~F., {Frasca}, A., {Alcal{\'a}}, J.~M., {et~al.} 2017, \aap, 605,
  A86, \dodoi{10.1051/0004-6361/201730807}

\bibitem[{{Manara} {et~al.}(2013){Manara}, {Testi}, {Rigliaco}, {Alcal{\'a}},
  {Natta}, {Stelzer}, {Biazzo}, {Covino}, {Covino}, {Cupani}, {D'Elia}, \&
  {Randich}}]{Manara2013Chr}
{Manara}, C.~F., {Testi}, L., {Rigliaco}, E., {et~al.} 2013, \aap, 551, A107,
  \dodoi{10.1051/0004-6361/201220921}

\bibitem[{{Marleau} {et~al.}(2024){Marleau}, {Aoyama}, {Hashimoto}, \&
  {Zhou}}]{Marleau2024TWA27B}
{Marleau}, G.-D., {Aoyama}, Y., {Hashimoto}, J., \& {Zhou}, Y. 2024, \apj, 964,
  70, \dodoi{10.3847/1538-4357/ad1ee9}

\bibitem[{{Muzerolle} {et~al.}(1998{\natexlab{a}}){Muzerolle}, {Calvet}, \&
  {Hartmann}}]{Muzerolle1998Model}
{Muzerolle}, J., {Calvet}, N., \& {Hartmann}, L. 1998{\natexlab{a}}, \apj, 492,
  743, \dodoi{10.1086/305069}

\bibitem[{{Muzerolle} {et~al.}(2001){Muzerolle}, {Calvet}, \&
  {Hartmann}}]{Muzerolle2001Model}
---. 2001, \apj, 550, 944, \dodoi{10.1086/319779}

\bibitem[{{Muzerolle} {et~al.}(1998{\natexlab{b}}){Muzerolle}, {Hartmann}, \&
  {Calvet}}]{Muzerolle1998Accretion}
{Muzerolle}, J., {Hartmann}, L., \& {Calvet}, N. 1998{\natexlab{b}}, \aj, 116,
  455, \dodoi{10.1086/300428}

\bibitem[{{Natta} {et~al.}(2004){Natta}, {Testi}, {Muzerolle}, {Randich},
  {Comer{\'o}n}, \& {Persi}}]{Natta2004YBD}
{Natta}, A., {Testi}, L., {Muzerolle}, J., {et~al.} 2004, \aap, 424, 603,
  \dodoi{10.1051/0004-6361:20040356}

\bibitem[{{Oke} {et~al.}(1995){Oke}, {Cohen}, {Carr}, {Cromer}, {Dingizian},
  {Harris}, {Labrecque}, {Lucinio}, {Schaal}, {Epps}, \&
  {Miller}}]{Oke1995LRIS}
{Oke}, J.~B., {Cohen}, J.~G., {Carr}, M., {et~al.} 1995, \pasp, 107, 375,
  \dodoi{10.1086/133562}

\bibitem[{{Ringqvist} {et~al.}(2023){Ringqvist}, {Viswanath}, {Aoyama},
  {Janson}, {Marleau}, \& {Brandeker}}]{Ringqvist2023Delorme1b}
{Ringqvist}, S.~C., {Viswanath}, G., {Aoyama}, Y., {et~al.} 2023, \aap, 669,
  L12, \dodoi{10.1051/0004-6361/202245424}

\bibitem[{{Santamar{\'\i}a-Miranda} {et~al.}(2018){Santamar{\'\i}a-Miranda},
  {C{\'a}ceres}, {Schreiber}, {Hardy}, {Bayo}, {Parsons}, {Gromadzki}, \&
  {Aguayo Villegas}}]{Santamaria-Miranda2018SR12c}
{Santamar{\'\i}a-Miranda}, A., {C{\'a}ceres}, C., {Schreiber}, M.~R., {et~al.}
  2018, Monthly Notices of the Royal Astronomical Society, 475, 2994,
  \dodoi{10.1093/mnras/stx3325}

\bibitem[{{Stamatellos} \& {Herczeg}(2015)}]{Stamatellos2015GI}
{Stamatellos}, D., \& {Herczeg}, G.~J. 2015, \mnras, 449, 3432,
  \dodoi{10.1093/mnras/stv526}

\bibitem[{{Tanigawa} \& {Watanabe}(2002)}]{tanigawa2002}
{Tanigawa}, T., \& {Watanabe}, S.-i. 2002, \apj, 580, 506,
  \dodoi{10.1086/343069}

\bibitem[{{Thanathibodee} {et~al.}(2019){Thanathibodee}, {Calvet}, {Bae},
  {Muzerolle}, \& {Hern{\'a}ndez}}]{Thanathibodee2019PDS70b}
{Thanathibodee}, T., {Calvet}, N., {Bae}, J., {Muzerolle}, J., \&
  {Hern{\'a}ndez}, R.~F. 2019, \apj, 885, 94, \dodoi{10.3847/1538-4357/ab44c1}

\bibitem[{{Thanathibodee} {et~al.}(2022){Thanathibodee}, {Calvet},
  {Hern{\'a}ndez}, {Mauc{\'o}}, \& {Brice{\~n}o}}]{Thanathibodee2022WTTS}
{Thanathibodee}, T., {Calvet}, N., {Hern{\'a}ndez}, J., {Mauc{\'o}}, K., \&
  {Brice{\~n}o}, C. 2022, \aj, 163, 74, \dodoi{10.3847/1538-3881/ac3ee6}

\bibitem[{{Theissen} {et~al.}(2018){Theissen}, {Burgasser}, {Bardalez
  Gagliuffi}, {Hardegree-Ullman}, {Gagn{\'e}}, {Schmidt}, \&
  {West}}]{Theissen2018-2m1115}
{Theissen}, C.~A., {Burgasser}, A.~J., {Bardalez Gagliuffi}, D.~C., {et~al.}
  2018, \apj, 853, 75, \dodoi{10.3847/1538-4357/aaa0cf}

\bibitem[{{Theissen} {et~al.}(2017){Theissen}, {West}, {Shippee}, {Burgasser},
  \& {Schmidt}}]{Theissen2017-2m1115}
{Theissen}, C.~A., {West}, A.~A., {Shippee}, G., {Burgasser}, A.~J., \&
  {Schmidt}, S.~J. 2017, \aj, 153, 92, \dodoi{10.3847/1538-3881/153/3/92}

\bibitem[{{Tody}(1986)}]{Tody1986IRAF}
{Tody}, D. 1986, in Society of Photo-Optical Instrumentation Engineers (SPIE)
  Conference Series, Vol. 627, Instrumentation in astronomy VI, ed. D.~L.
  {Crawford}, 733, \dodoi{10.1117/12.968154}

\bibitem[{{van der Walt} {et~al.}(2011){van der Walt}, {Colbert}, \&
  {Varoquaux}}]{VanDerWalt2011}
{van der Walt}, S., {Colbert}, S.~C., \& {Varoquaux}, G. 2011, Computing in
  Science and Engineering, 13, 22, \dodoi{10.1109/MCSE.2011.37}

\bibitem[{{Venuti} {et~al.}(2019){Venuti}, {Stelzer}, {Alcal{\'a}}, {Manara},
  {Frasca}, {Jayawardhana}, {Antoniucci}, {Argiroffi}, {Natta}, {Nisini},
  {Randich}, \& {Scholz}}]{Venuti2019TWA}
{Venuti}, L., {Stelzer}, B., {Alcal{\'a}}, J.~M., {et~al.} 2019, \aap, 632,
  A46, \dodoi{10.1051/0004-6361/201935745}

\bibitem[{{Vernet} {et~al.}(2011){Vernet}, {Dekker}, {D'Odorico}, {Kaper},
  {Kjaergaard}, {Hammer}, {Randich}, {Zerbi}, {Groot}, {Hjorth}, {Guinouard},
  {Navarro}, {Adolfse}, {Albers}, {Amans}, {Andersen}, {Andersen}, {Binetruy},
  {Bristow}, {Castillo}, {Chemla}, {Christensen}, {Conconi}, {Conzelmann},
  {Dam}, {de Caprio}, {de Ugarte Postigo}, {Delabre}, {di Marcantonio},
  {Downing}, {Elswijk}, {Finger}, {Fischer}, {Flores}, {Fran{\c{c}}ois},
  {Goldoni}, {Guglielmi}, {Haigron}, {Hanenburg}, {Hendriks}, {Horrobin},
  {Horville}, {Jessen}, {Kerber}, {Kern}, {Kiekebusch}, {Kleszcz}, {Klougart},
  {Kragt}, {Larsen}, {Lizon}, {Lucuix}, {Mainieri}, {Manuputy}, {Martayan},
  {Mason}, {Mazzoleni}, {Michaelsen}, {Modigliani}, {Moehler}, {M{\o}ller},
  {Norup S{\o}rensen}, {N{\o}rregaard}, {P{\'e}roux}, {Patat}, {Pena}, {Pragt},
  {Reinero}, {Rigal}, {Riva}, {Roelfsema}, {Royer}, {Sacco}, {Santin},
  {Schoenmaker}, {Spano}, {Sweers}, {Ter Horst}, {Tintori}, {Tromp}, {van
  Dael}, {van der Vliet}, {Venema}, {Vidali}, {Vinther}, {Vola}, {Winters},
  {Wistisen}, {Wulterkens}, \& {Zacchei}}]{Vernet2011Xshooter}
{Vernet}, J., {Dekker}, H., {D'Odorico}, S., {et~al.} 2011, \aap, 536, A105,
  \dodoi{10.1051/0004-6361/201117752}

\bibitem[{{White} \& {Basri}(2003)}]{White2003EW}
{White}, R.~J., \& {Basri}, G. 2003, \apj, 582, 1109, \dodoi{10.1086/344673}

\bibitem[{{Wilson} {et~al.}(2022){Wilson}, {Matt}, {Harries}, \&
  {Herczeg}}]{Wilson2022Model}
{Wilson}, T.~J.~G., {Matt}, S., {Harries}, T.~J., \& {Herczeg}, G.~J. 2022,
  \mnras, 514, 2162, \dodoi{10.1093/mnras/stac1397}

\bibitem[{{York} {et~al.}(2000){York}, {Adelman}, {Anderson}, {Anderson},
  {Annis}, {Bahcall}, {Bakken}, {Barkhouser}, {Bastian}, {Berman}, {Boroski},
  {Bracker}, {Briegel}, {Briggs}, {Brinkmann}, {Brunner}, {Burles}, {Carey},
  {Carr}, {Castander}, {Chen}, {Colestock}, {Connolly}, {Crocker}, {Csabai},
  {Czarapata}, {Davis}, {Doi}, {Dombeck}, {Eisenstein}, {Ellman}, {Elms},
  {Evans}, {Fan}, {Federwitz}, {Fiscelli}, {Friedman}, {Frieman}, {Fukugita},
  {Gillespie}, {Gunn}, {Gurbani}, {de Haas}, {Haldeman}, {Harris}, {Hayes},
  {Heckman}, {Hennessy}, {Hindsley}, {Holm}, {Holmgren}, {Huang}, {Hull},
  {Husby}, {Ichikawa}, {Ichikawa}, {Ivezi{\'c}}, {Kent}, {Kim}, {Kinney},
  {Klaene}, {Kleinman}, {Kleinman}, {Knapp}, {Korienek}, {Kron}, {Kunszt},
  {Lamb}, {Lee}, {Leger}, {Limmongkol}, {Lindenmeyer}, {Long}, {Loomis},
  {Loveday}, {Lucinio}, {Lupton}, {MacKinnon}, {Mannery}, {Mantsch}, {Margon},
  {McGehee}, {McKay}, {Meiksin}, {Merelli}, {Monet}, {Munn}, {Narayanan},
  {Nash}, {Neilsen}, {Neswold}, {Newberg}, {Nichol}, {Nicinski}, {Nonino},
  {Okada}, {Okamura}, {Ostriker}, {Owen}, {Pauls}, {Peoples}, {Peterson},
  {Petravick}, {Pier}, {Pope}, {Pordes}, {Prosapio}, {Rechenmacher}, {Quinn},
  {Richards}, {Richmond}, {Rivetta}, {Rockosi}, {Ruthmansdorfer}, {Sandford},
  {Schlegel}, {Schneider}, {Sekiguchi}, {Sergey}, {Shimasaku}, {Siegmund},
  {Smee}, {Smith}, {Snedden}, {Stone}, {Stoughton}, {Strauss}, {Stubbs},
  {SubbaRao}, {Szalay}, {Szapudi}, {Szokoly}, {Thakar}, {Tremonti}, {Tucker},
  {Uomoto}, {Vanden Berk}, {Vogeley}, {Waddell}, {Wang}, {Watanabe},
  {Weinberg}, {Yanny}, {Yasuda}, \& {SDSS Collaboration}}]{York2000SDSS}
{York}, D.~G., {Adelman}, J., {Anderson}, John~E., J., {et~al.} 2000, \aj, 120,
  1579, \dodoi{10.1086/301513}

\bibitem[{{Zhou} {et~al.}(2014){Zhou}, {Herczeg}, {Kraus}, {Metchev}, \&
  {Cruz}}]{zhou14}
{Zhou}, Y., {Herczeg}, G.~J., {Kraus}, A.~L., {Metchev}, S., \& {Cruz}, K.~L.
  2014, The Astrophysical Journal, 783, L17,
  \dodoi{10.1088/2041-8205/783/1/L17}

\bibitem[{{Zhou} {et~al.}(2021){Zhou}, {Bowler}, {Wagner}, {Schneider}, {Apai},
  {Kraus}, {Close}, {Herczeg}, \& {Fang}}]{Zhou2021PDS70}
{Zhou}, Y., {Bowler}, B.~P., {Wagner}, K.~R., {et~al.} 2021, \aj, 161, 244,
  \dodoi{10.3847/1538-3881/abeb7a}

\end{thebibliography}
\bibliographystyle{aasjournal}

\end{document}